\documentclass[prr,aps,twocolumn,twoside,superscriptaddress,floats,floatfix]{revtex4}
\usepackage{amssymb,amsmath}
\usepackage[pdftex]{graphicx}
\usepackage{array,multirow}
\usepackage{hyperref}
\usepackage[usenames,dvipsnames]{color}
\usepackage{caption}
\usepackage{pifont}
\usepackage{stix}
\usepackage{comment}
\usepackage{verbatim}
\usepackage{soul}
\usepackage{listings}
\usepackage{boldline}
\usepackage{float}
\usepackage{array}
\begin{document}
\title{ Spiral- and scroll-wave dynamics in mathematical models for canine and human 
ventricular tissue with varying Potassium and Calcium currents}
\author{K. V. Rajany}
\email{rajanyk@iisc.ac.in}
\affiliation{Centre for Condensed Matter Theory, Department of Physics, Indian
Institute of Science, Bangalore 560012, India. } 
\altaffiliation[Present address:]{S.A.R.B.T.M. Govt. College, Koyilandy, Calicut, India}
\author{Alok Ranjan Nayak}
\affiliation{International Institute of Information Technology, Bhubaneswar.}
\author{Rupamanjari Majumder}
\affiliation{Department of Fluid Dynamics, Pattern Formation and Biocomplexity
Max Planck Institute for Dynamics and Self-Organization.}
\author{Rahul Pandit}
\email{rahul@iisc.ac.in}
\altaffiliation[\\ also at~]{Jawaharlal Nehru Centre For Advanced
Scientific Research, Jakkur, Bangalore, India. }
\affiliation{Centre for Condensed Matter Theory, Department of Physics, Indian
Institute of Science, Bangalore 560012, India. } 
\begin{abstract}
We conduct a systematic, direct-numerical-simulation (DNS) study, in
mathematical models for ventricular tissue, of the dependence of
spiral- and scroll-wave dynamics on $G_{Kr}$, the maximal conductance
of the delayed rectifier Potassium current ($I_{Kr}$) channel, and the
parameter $\gamma_{Cao}$, which determines the magnitude and shape of
the current $I_{CaL}$ for the L-type calcium-current channel, in both
square and anatomically realistic, whole-ventricle simulation domains.
We study canine and human models. In the former, we use a
canine-ventricular geometry, with fiber-orientation details, obtained
from diffusion-tensor-magnetic-resonance-imaging (DTMRI) data; and we
employ the physiologically realistic Hund-Rudy-Dynamic (HRD) model
for a canine ventricular myocyte. To focus on the dependence of spiral-
and scroll-wave dynamics on $G_{Kr}$ and $\gamma_{Cao}$, we restrict
ourselves to an HRD-model parameter regime, which does not produce
spiral- and scroll-wave instabilities because of other, well-studied
causes like a very sharp action-potential-duration-restitution (APDR)
curve or early after 
depolarizations (EADs) at the single-cell level.
We find that spiral- or scroll-wave dynamics are affected predominantly
by a simultaneous change in $I_{CaL}$ and $I_{Kr}$, rather than by a
change in any one of these currents; other currents do not have such a
large effect on these wave dynamics in this parameter regime of the
HRD model. In particular, we examine spiral-wave dynamics for ten
different values of $G_{Kr}$ and ten different values of $\gamma_{Cao}$
in our 2D DNSs. For our 3D DNSs in an anatomically realistic domain, we
chose 16 parameter sets. In the parameter regime we begin with, the
system displays broken spiral or scroll states with S1-S2 initial
conditions (see below). We show that, by simultaneously increasing
$G_{Kr}$ and reducing $\gamma_{Cao}$, we can get to a parameter regime
in which the system displays single, stable rotating spirals or scroll
waves. We obtain stability diagrams (or phase diagrams) in the $G_{Kr}
- \gamma_{Cao}$ plane; and we find that these diagrams are
significantly different in our 2D and 3D studies. In the 3D case, the
geometry of the domain itself supports the confinement of the scroll
waves and makes them more stable compared to their spiral-wave
counterparts in our flat, 2D simulation domain. Thus, a combination of
functional and geometrical mechanisms produce different dynamics for 3D
scroll waves and their 2D spiral-wave counterparts. In particular, the
former do not break easily because, in an anatomically realistic
ventricular geometry, they are not easily absorbed at boundaries, nor
do they break near boundaries. We have also carried out a comparison of
our HRD results with their counterparts for the human-ventricular
TP06 model; and we have found important differences between wave
dynamics in these two models. The region in parameter space, where we
obtain broken spiral or scroll waves in the HRD model is the region
of stable rotating waves in the TP06 model; the default parameter
values produce broken waves in the HRD model, but stable scrolls in
the TP06 model. In both these models, to make a transition, (most
simply, from broken-wave to stable-scroll states) we must
simultaneously increase $I_{Kr}$ and decrease $I_{CaL}$; a modification
of only one of these currents is not enough to effect this transition.
Furthermore, the converse, i.e., an increase in $I_{CaL}$ along with a
decrease in $I_{Kr}$ does not yield any interesting dynamical
transitions in the HRD model, for, in this range of currents, this
model does not sustain spiral or scroll waves or broken waves. 
\end{abstract}
\maketitle
\date{\today}
\section{Introduction}

Studies of mathematical models of cardiac myocytes and cardiac tissue play an
important role in understanding the complex mechanisms that underlie cardiac
arrhythmias, which are a major cause of death. These arrhythmias are believed
to be associated with reentrant waves of electrical activation in cardiac
tissue; specifically, rotating spiral or scroll waves are associated with
ventricular tachycardia (VT) and the breaking of such waves is associated with
ventricular fibrillation (VF). Understanding the detailed ionic mechanisms
leading to spiral or scroll waves and the response of these wave dynamics to
the changes in various ionic mechanisms is still a difficult task, because we
must account for the large number of ion channels, ion pumps, and intracellular
mechanisms that are involved in producing the action potential (AP).
Computational tools are becoming more and more useful in these studies (~\cite{davidenko93,clayton08,
clayton11, trayanova11, cherry08, sneyd, reviewroyal, review1, rpm1, shajahan1}) because
they allow us increased control and flexibility in handling each parameter and, 
hence, the associated ionic current; such control is rarely feasible in
experiments. A number of studies have been conducted on the mechanisms of
ion-channel kinetics and their dependence on various channel parameters or the
AP morphology.
 But there are a  few detailed studies of how
the current-channel parameters directly affect scroll waves, at the whole-heart
level in three dimensions (3D). Furthermore, experiments on mammalian hearts are challenging because of the
difficulty in the visualization of waves of electrical activation below the tissue surface~\cite{rpm1,shajahan1,rpm2,rpm3,alok2,alok3,ikeda,limmaskara,shajahan2,shajahan3}; for recent advances in such visualization, we refer the reader to Refs.~\cite{christoph, Grondin19}.


Canine hearts are often used in studies of the mechanisms of cardiac
arrhythmias because their size and physiology are comparable to those of human
hearts. Furthermore, detailed mathematical models for canine cardiac tissue are
now available; these models incorporate important electrophysiological details,
including intracellular calcium dynamics. In our work, we use a detailed
mathematical model for canine ventricular myocytes, namely, the
Hund-Rudy-Dynamic (HRD) model~\cite{hrdmodel,hrd-supplementary}. We use a canine ventricular geometry, with
fiber-orientation details, which have been obtained in earlier studies that use
diffusion-tensor magnetic-resonance imaging (DTMRI). These DTMRI data have been
made available for academic use at the CMISS site (https://www.cmiss.org/).
Most of the studies on scroll-wave dynamics have been carried out on models
that are not as detailed as the HRD model; e.g., many studies use the Luo-Rudy
model for guinea pigs~\cite{luo_rudy}. Thus, our work goes significantly beyond
such earlier studies. 

In particular, we carry out a systematic, \textit{in silico}
direct-numerical-simulation (DNS) of spiral and scroll waves in the HRD
mathematical model for canine ventricular tissue. We explore the dependence of
spiral- and scroll-wave dynamics on $G_{Kr}$, the maximal conductance of the
delayed rectifier Potassium current ($I_{Kr}$) channel, and the parameter
$\gamma_{Cao}$, which determines the magnitude and shape of the current
$I_{CaL}$ for the L-type calcium-current channel, in both square and
anatomically realistic, whole-ventricle simulation domains. We focus on the
dependence of spiral- and scroll-wave dynamics on $G_{Kr}$ and $\gamma_{Cao}$;
therefore, we limit ourselves to a parameter regime, in which the HRD model
does not display spiral- and scroll-wave instabilities arising from other
well-explored causes, like a very sharp action-potential-duration-restitution
(APDR)~\cite{fenton:02, garfinkel:00, koller:98} curve or early after depolarizations (EADs) at the single-cell level.
We find that spiral- or scroll-wave dynamics are affected predominantly by a
simultaneous change in $I_{CaL}$ and $I_{Kr}$, rather than by a change in any
one of these currents; other currents do not display such a large effect on
these wave dynamics in this parameter regime. To carry out a systematic study,
we examine spiral-wave dynamics for ten different values of $G_{Kr}$ and ten
different values of $\gamma_{Cao}$ in our 2D DNSs. For our 3D DNSs in an
anatomically realistic domain, we choose 16 parameter sets. In the parameter
regime we begin with, the system displays broken spiral or scroll states with
the S1-S2 initial conditions (see below). We show that, by simultaneously
increasing $G_{Kr}$ and reducing $\gamma_{Cao}$, we reach a parameter regime in
which the system displays a single, stable rotating spiral or scroll wave. We
obtain stability diagrams (or phase diagrams) in the $G_{Kr} - \gamma_{Cao}$
plane; and we find that these diagrams are significantly different in our 2D
and 3D studies. In the 3D case, the geometry of the domain itself supports the
confinement of the scroll waves and makes them more stable compared to their
spiral-wave counterparts in our flat, 2D simulation domain. Thus, a combination
of functional and geometrical mechanisms produce different dynamics for 3D
scroll waves and their 2D spiral-wave counterparts: The former do not break
easily because, in an anatomically realistic geometry, they are not easily
absorbed at boundaries; nor do they break near boundaries. 

We have mentioned above that canine hearts are considered to be similar to
human hearts in both shape and electrophysiology. It is important to explore
this similarity. We begin such and exploration by comparing our HRD-model
results with their counterparts for the human-ventricular TP06 mathematical
model~\cite{tp06}. We use a human-ventricular geometry with fiber rotation; to
obtain the coordinates in the human-ventricular geometry we use
Ref.~\cite{humangeomdata}. By carrying out \textit{in silico} DNSs of spiral
and scroll waves in this TP06 model, we find important differences between wave
dynamics in these two models. The region in parameter space, where we obtain
broken spiral or scroll waves in the HRD model is the region of stable rotating
waves in the TP06 model; the default parameter values produce broken waves in
the HRD model, but stable scrolls in the TP06 model. However, in both these
models, to make a transition (most simply, from broken-wave to stable-scroll
states), we must simultaneously increase $I_{Kr}$ and decrease $I_{CaL}$; a
modification of only one of these currents is not enough to effect this
transition. Furthermore, the converse, i.e., an increase in $I_{CaL}$ along
with a decrease in $I_{Kr}$ does not yield any interesting dynamical
transitions in the HRD model, for, in this range of currents, this model does
not sustain unbroken or broken waves. 


{ In the Supplementary Material~\cite{supplementary}, we
describe the models we use for our DNSs, namely, the HRD model, for a
canine-ventricular myocyte, and the TP06 model, for a human-ventricular
myocyte. We give a description of the two currents which are the subject of
this study in each models. We also describe the anatomically realistic geometry
and the numerical methods that we use to study scroll dynamics}.

The remaining part of this paper is organized as follows.
{Section \label{c3-sec:methods} describes briefly the DNSs we
have conducted, the details of which are given in the Supplementary
Material~\cite{supplementary}. } Sections II and III are devoted to our
results, which are presented in two parts, the first for the HRD model and the
second for the TP06 model.  Each part has {two} subsections,
that are devoted, respectively, to our results for 2D tissue, and 3D
anatomically realistic domains; we compare our results from HRD and TP06
models.  {We describe the results of our cellular-level studies
in the Supplementary Material~\cite{supplementary}. We also present the
variation of the two important currents, which we focus on in this study (as we
change model parameters), along with the corresponding APDR curves, for both
HRD and TP06 models}.  Section III contains a discussion of our results and
conclusions. 


\section{Models and Numerical Methods} \subsection{Canine Ventricular (HRD
Model) Simulations} \label{c3-subsec:CVS}

We have used the physiologically detailed HRD mathematical model for canine
ventricular tissue. This is a dynamic model that reproduces the experimentally
measured action potential(AP) and Calcium-current regulation over a wide range
of the pacing frequency. The HRD model incorporates a total of 15 ionic
currents:
\begin{gather*}
I_{Na}, I_{NaL}, I_{CaL}, I_{NaCa}, I_{NaK}, I_{Ks}, I_{Kr}, I_{to1}, I_{to2},\\
	I_{K1}, I_{Kp}, I_{Cab}, I_{Clb}, I_{pCa}, I_{rel}.
\end{gather*}
This model uses $21$ gating variables, namely,
\begin{gather*}
 H, m, J, d, f, f_2, f_{ca}, \\
f_{ca2}, p, r, x_r, x_{s1}, x_{s2}, y,\\
ydv, ydv2, AA, m_L, h_L, r_o, r_i 
\end{gather*}
and the following $8$ ionic concentrations: \[Ca_i, Na_i, Cl_i, K_i, Ca_{ss},
Ca_{jsr}, Ca_{nsr}, CaMK_{trap}\].  The details of the model and complete
equations are given in the Supplementary Material~\cite{supplementary}.
In this model we calculate the transmembrane potential $V$ as a
dynamical function of the above mentioned currents, concentrations, and gating
variables. 
{We give tables with (a) a list of the currents in the HRD model
and their descriptions and (b) ionic concentrations in the
Supplementary Material~\cite{supplementary}.}

\begin{figure*}
\includegraphics[width=1\linewidth]{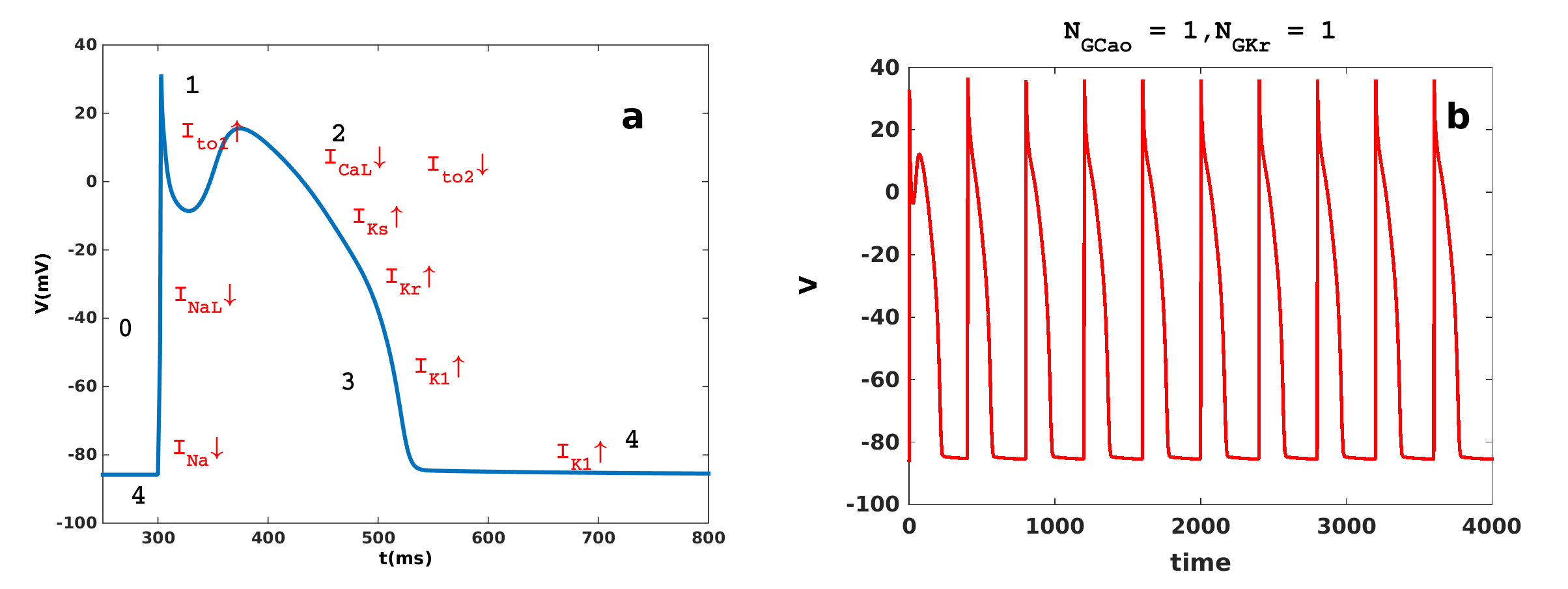}
\caption[Action potentials for the canine-myocyte HRD model]{(a) Plot of the transmembrane potential $V$ versus time, for the HRD
canine-ventricular model cell, showing the shape of the action potential (AP) 
and the major currents responsible for each phase of this AP. (b) Plots 
of these APs versus time for repeated pacing of a single cell, with fixed (representative)
values of $G_{Kr}$ and of $\gamma_{Cao}$ (see the text and compare with 
Fig.~\ref{c3-aphuman} for the human-myocyte TP06 model).}
\label{c3-apdog}
\end{figure*}
\begin{figure*}
\includegraphics[width=1\linewidth]{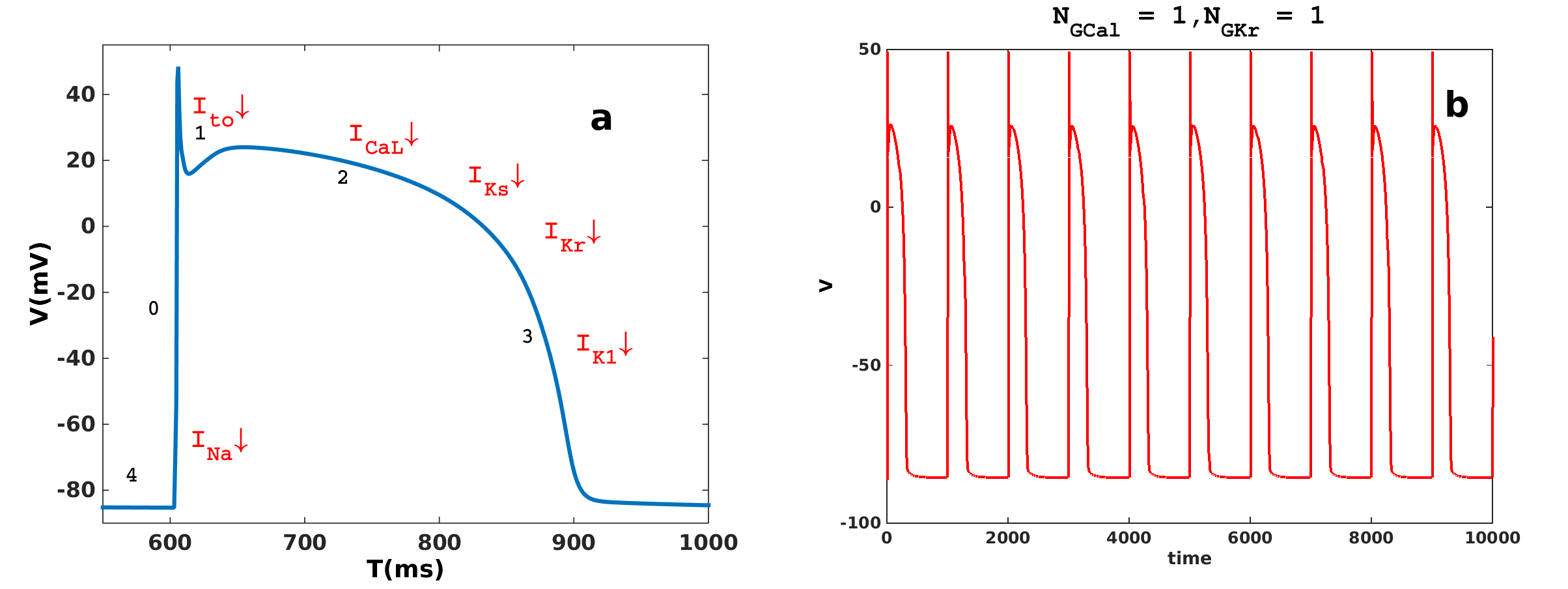}
\caption[Action potentials for the human-myocyte TP06 model]{(a) Plot of the transmembrane potential $V$ versus time, for the TP06
human-ventricular model cell, showing the shape of the action potential (AP) 
and the major currents responsible for each phase of this AP. (b) Plots 
of these APs versus time for repeated pacing of a single cell, with fixed (representative)
values of $G_{Kr}$ and of $G_{CaL}$ (see text and compare with Fig.~\ref{c3-apdog}
for the canine-myocyte HRD model).}
\label{c3-aphuman}
\end{figure*}

We depict in Fig.~\ref{c3-apdog} the shape of the action potential AP  for the
HRD canine-myocyte model, showing its different phases along with the major
currents responsible for each stage. This figure also shows, for a
representative set of values for $G_{Kr}$ and $\gamma_{Cao}$, the APs that we
obtain with continuous pacing of the myocyte. We show for comparison, in
Fig.~\ref{c3-aphuman}, similar plots for the TP06 human-myocyte model.

Our 2D simulation domain, for the HRD model, is a square tissue with size
$8.32$ cm $\times 8.32$ cm. For our 3D simulation we use the processed
Diffusion-Tensor Magnetic-Resonance Imaging (DTMRI) data for the
canine-ventricular anatomy, which is freely available for academic purposes at
	the CMISS website\\ (https://www.cmiss.org/),{ the details 
	of which are given in the Supplementary Material~\cite{supplementary}.}

Fig.~\ref{c3-anat_geom} shows the geometry of the domain in which
we investigate scroll-wave dynamics. 

{We describe the S1-S2 proptocol, which we use to produce the initial 
            scroll waves in the Supplementary Material~\cite{supplementary}.}
\begin{figure}
	\includegraphics[width=0.8\linewidth]{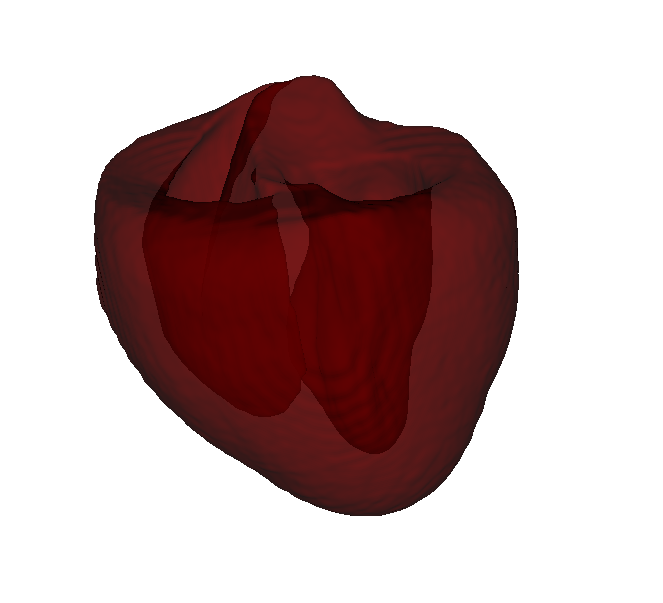}

\caption[Anatomical geometry of the canine heart]{The anatomical geometry that we have created for our
simulations by using DTMRI data (see text) for the coordinate mesh for
the canine heart. 
Our simulation domain is the lower part of the heart that comprises the ventricles and 
the septum.
}
	\label{c3-anat_geom}
\end{figure}
\begin{figure}
		\includegraphics[width=0.8\linewidth]{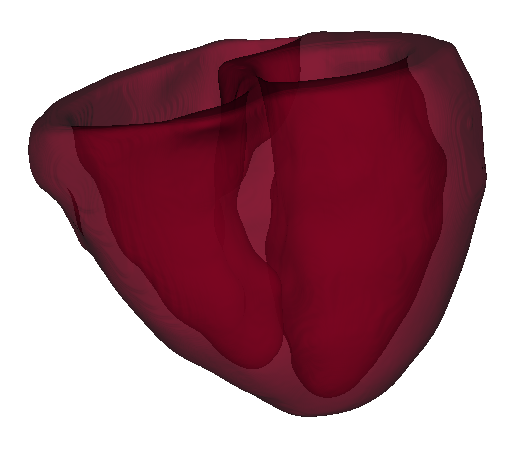}
\caption[Anatomical geometry of the human heart]{The anatomical geometry that we have created for our simulations 
of the TP06 human-ventricular model. We use the data for the coordinate mesh of the 
human heart that has been made available at Ref.~\cite{humangeomdata}. Our 
simulation domain is the lower part of the heart that comprises the ventricles and 
the septum.}
\label{c3-hanat_geom}
\end{figure}

For the HRD model, we investigate the effect on spiral- and scroll-wave dynamics 
of the following two currents. 

(1) The L-Type Calcium current that is given by the equations
\begin{eqnarray}
I_{CaL} & = &d^{p}\cdot f_{ca}\cdot f_{ca2} \cdot f \cdot {\bar{I}}_{CaL}, \\
{\bar{I}}_{CaL} & = &P_{Ca} \cdot z^{2}_{Ca}. \frac{(V_m-15. 0). F^2}{RT}.\\
& &  \frac{\gamma_{Cai} . [Ca]_{ss}\cdot \exp(\frac{z_{Ca}(V_m-15. 0)F}{RT})-\gamma_{Cao}\cdot [Ca]_o}{\exp(\frac{z_{Ca}(V_m-15. 0)F}{RT})-1}, \\
\label{c3-icaleqn}
\end{eqnarray}
where $d$, $p$, $f_{ca}$, $f_{ca2}$, and $f$ are gating variables; 
$P_{Ca}$ { is the membrane permeability to the $Ca$ ion, in units of
$cm/s$, and $z_{Ca}$ is the valence of the $Ca$  ion.}
The parameters $\gamma_{Cai}$ and $\gamma_{Cao}$
determine the magnitude and shape of the current $I_{CaL}$.
{They are the activity coefficient of the $Ca$ ion.} 
We find that
$\gamma_{Cao}$ has a stronger influence on the dynamics of the system than
does $\gamma_{Cai}$. Therefore, we have chosen to study how the variation of 
$\gamma_{Cao}$, over a wide range, affects spiral- and scroll-wave dynamics in the 
HRD model. 

(2) The delayed rectifier Potassium current is
\begin{eqnarray}
I_{Kr}&=&\bar{G}_{Kr}\cdot X_r \cdot R_{Kr} \cdot (V_m-E_{Kr}) , \\
\bar{G}_{Kr}&=& G_{Kr} \cdot \sqrt{\frac{[K^+]_o}{5. 4}} , \\
\label{c3-ikreqn}
\end{eqnarray}
{where $\bar{G}_{Kr}$ is the maximal conductance in units of $mS/\mu F$.}
 { $[K^+]_o$ is the extracellular concentration of $K^+$ ion in mmol/L .}
We study spiral- and scroll-wave dynamics in the HRD model for a wide range of 
values of $G_{Kr}$. The currents are given in units of $\mu A/ \mu F$ .

In Eqs. \ref{c3-icaleqn} and ~\ref{c3-ikreqn}, $G_{Kr}= 0.01385423$ and
$\gamma_{Cao}=0.341$; henceforth, we refer to these values as $GKR$ and $GCAO$,
respectively.  We investigate the dynamics of spiral waves in 2D tissue for 100
different cases, by varying $G_{Kr}$ as $GKR \times N_{GKr}$, where $N_{GKr}=
1, 2, ...10$, and, simultaneously, varying $\gamma_{Cao}$ as $GCAO/N_{GCao}$,
where $N_{GCao}=1, 2...10$. 

\subsection{Human-Ventricular (TP06 Model) Simulations}
\label{c3-subsec:HVS}


For our DNSs of human-ventricular tissue, we use the TP06(\cite{tp06}) model, 
which is a modified version of the TNNP model for human-ventricular
cells~\cite{tnnp04}; this incorporates a total of $12$ ionic currents. The
intracellular calcium handling in these models is not as detailed as it is in
the HRD model. 

The TP06 model uses 19 variables: (a) 1 for the transmembrane potential $V_m$, (b) 13
for ion-channel gates, namely, $m$, $h$, $j$, $d$, $f$, $f_2$, 
$f_{cass}$, $r$, $s$, $x_s$, $x_{r1}$, $x_{r2}$, and $\bar R$, 
and (c) 5 for intracellular, ion-concentration dynamics, namely, 
$Na_i$, $K_i$, $Ca_i$, $Ca_{sr}$, and $Ca_{ss}$. 


As we have mentioned above, Fig.~\ref{c3-aphuman} gives the shape of the AP for
this TP06 model; it depicts different, phases along with the major currents
responsible for each phase; this figures also shows, for representative values
of $G_{Kr}=GKR$ and $G_{CaL}=GCAL$, the APs with continuous pacing of the
myocyte. 


In Fig.~\ref{c3-hanat_geom} we have shown the anatomical geometry that we have created
for our simulations by using the DTMRI data for the coordinate mesh of the human
heart. This is the lower part of the heart that contains the ventricles and 
the septum.

The L-Type calcium current in the TP06 model is described by the following
equations:
\begin{eqnarray}
I_{CaL} &=& G_{CaL}. d. f. f_{Ca}\cdot4\cdot{\frac {(V-15)F^2}{RT}}\cdot \\
&& {\frac {0.25 Ca_{SS} \exp({\frac {2(V-15)F}{RT}})-Ca_o}{\exp(\frac{2(V-F)F}{RT})-1}},
\label{c3-hicaleqn}
\end{eqnarray}
where $d$, $f_{ca}$, and $f$ are gating variables. The parameter $G_{CaL}$
determines the magnitude and shape of the current $I_{CaL}$. As in our work on
the HRD model, we study how the variation of $G_{CaL}$, over a wide range, 
affects spiral- and scroll-wave dynamics in the TP06 model. Note that the
variation of $G_{CaL}$ in the TP06 model corresponds to the variation of
$\gamma_{Cao}$ in the HRD model. 

In the TP06 model, the delayed rectifier Potassium current is 
\begin{eqnarray}
I_{Kr} &=& G_{Kr}{\sqrt{\frac {{K_o}}{5.4}}}.x_{r1}.x_{r2}.(V-E_{K}). 
\label{c3-hikreqn}
\end{eqnarray}
In our DNSs, we investigate $20$ different cases in 2D tissue:
$G_{Kr}$ is taken as $G_{Kr}=GKR\times N_{GKr}$, where $GKR=0.153$, and 
$N_{GKr}=1, 3, . . 7$. Along with this, $G_{CaL}$ is varied as
$G_{CaL} = GCAL/N_{GCaL}$, where $GCAL=0.0000398$ and $N_{GCaL}=1, 3, . . 9$. 
$GKR$ is the original value of $G_{Kr}$, as it appears in the model described by 
Eq.~\ref{c3-hikreqn}; likewise, $GCAL$ is the original value of $G_{CaL}$ as it appears in 
the model described by Eq.~\ref{c3-hicaleqn}. 

For our DNSs in the 3D anatomical geometry, we investigate 20 different cases
as follows: $G_{Kr}=GKR\times N_{GKr}$, $N_{GKr}=1, 3, ...9$; and $ \\ G_{CaL} =
GCAL/N_{GCaL}$, $N_{GCaL}=1, 3, ...7$.

\section{Results}
\label{c3-sec:results}

{We present results from our simulations of the HRD
model, at the cell level, in the Supplementary Material~\cite{supplementary}. 
The variations of the AP morphologies and APDR curves, for all the parameter values in
our DNSs, are presented along with the plots of the currents $I_{Kr}$ and $
I_{CaL}$. In this Section,} we give results from our studies in 2D and then in
3D for the HRD model. Next we present the results of our simulations for the
TP06 model.

It has been observed previously that the action potential duration restitution
(APDR) is a crucial factor, which determines whether scroll waves or broken
scroll waves develop in cardiac tissue~\cite{fenton:02, garfinkel:00, koller:98, APDR1, APDR2}. The APDR is the
shortening of the action potential duration(APD) as we increase the pacing
frequency. A sharp APDR with a slope $>1$ in the APDR curve usually gives 
rise to a chaotic, broken-wave pattern. Another crucial determinant for the
break up of spiral or scroll waves is early after depolarization (EAD), a
premature re-excitation of the recovering tissue~\cite{EAD1, Soling1, Soling2}.
We examine a parameter region in the HRD model where we see neither EADs nor
a sharp APDR curve (its slope is always $<1$). 
\subsection{2D Results}
\label{c3-dog2d}

The conduction velocity $cv$ of a plane wave, passing from one end to the other
end in our 2D simulation domain, is the same for all the cases we study; we
find $cv = 6.7002 m/s$. By contrast, the wavelength $\lambda$ of the plane
wave varies for each case; we define the \textit {wavelength} to be the
distance between the excited front and the $90\%$ recovered back end of the
propagating plane wave of the transmembrane potential $V$ (before the
initiation of the spiral wave). As we have noted above, we can also use the
formula $\lambda= cv\times APD$, where APD is the action potential duration.
We measure $\lambda$ for each of our parameter sets. The spiral wave, which we
use as initial condition for our 2D DNSs, is created by using the S1-S2 protocol,
{which we describe in the Supplementary Material,~\cite{supplementary},
where we give representative pseudocolor plots of $V$ for plane-wave and spiral-wave 
initial conditions.}
\begin{figure}
\includegraphics[width=0.9\linewidth]{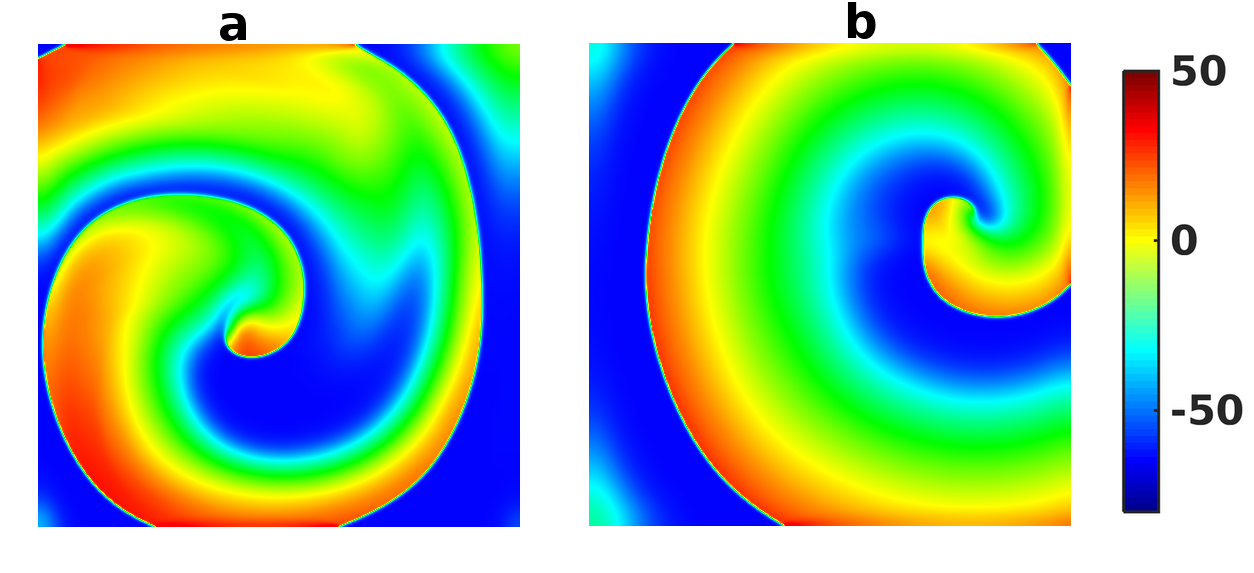}
\caption[Two kinds of spiral-wave arms formed in our 2D tissue domain for HRD model]
{Pseudocolor plots of the transmembrane potential $V$ showing the two major 
types of spiral arms we observe in our 2D HRD-model simulations: (a) a spiral with a
nonuniform wave width in different regions, with much thinner arms than in the center;
and (b) a spiral with an almost uniform arm width. For the complete spatiotemporal 
evolution see the Videos S33 and S34 in the Supplementary Material~\cite{supplementary}.} 
\label{c3-spiral_arm}
\end{figure}
%
\begin{figure}
\includegraphics[width=0.9\linewidth]{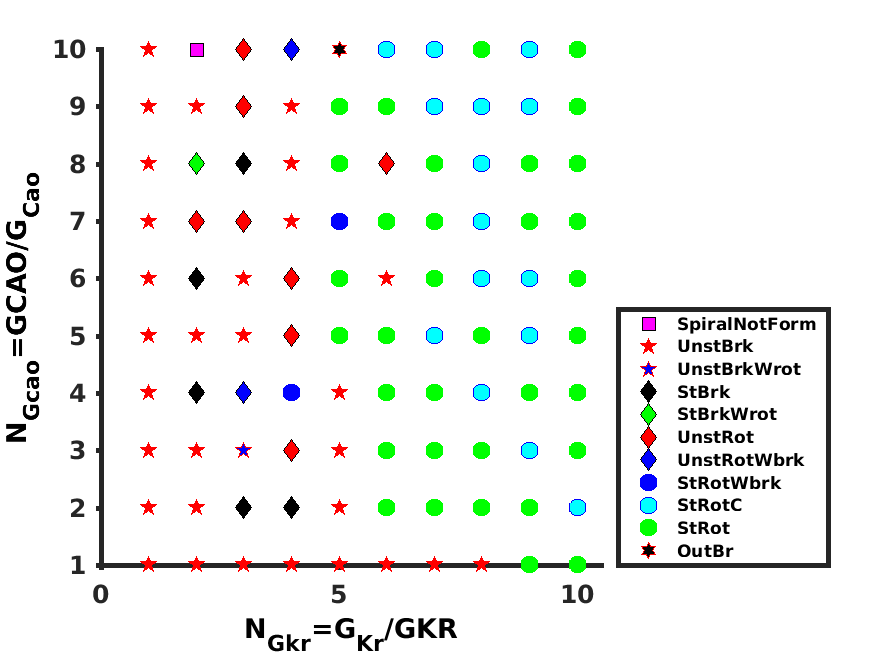}
	\caption[Phase diagram (or stability diagram) for all types of spiral-wave
dynamics observed in our 2D domain for the HRD model]{Phase diagram (or stability diagram) for the types of spiral-wave 
dynamics in our 2D domain for the HRD model.
.The markers show the exact type of wave dynamics by using different shapes and colors. 
In the yellow region we obtain rotating spirals (${\mdblkcircle}$ with different colors);
in the blue region we get broken spirals (\ding{72}); and in the green region we see 
a mixed behavior, with transitions from unbroken spirals to states with broken spirals
and vice-versa (diamond, ${\mdblkdiamond}$). In each region there is a statistically 
stable regime (waves remain in the medium without decaying) and an unstable regime 
(waves move away and disappear completely). In some cases, no spiral is formed
(e.g., when $N_{GKr}=2, N_{GCao}=10$, denoted by a magenta square $\mdblksquare$).
For $N_{GKr}=5$, $\gamma_{Cao}=10$, we see a far-field-break-up phenomenon 
(black-faced hexagon\ding{86}). The acronyms we use are as follows:
SpiralNotForm- no spiral; UnstBrk - unstable break-up; UnstBrkWrot - unstable break-up 
state with rotating spiral states appearing temporarily; StBrk - stable break-up; 
StBrkWrot - stable break-up state with rotating spiral states appearing temporarily; 
UnstRot - unstable rotating spiral; UnstRotWbrk - unstable rotating spiral state with 
intermittent temporary spiral-break-up states; StRotWbrk - stable rotating spiral state 
with intermittent temporary spiral-break-up states; StRotC - stable rotating spiral 
state with temporary tiny break-up in the center; StRot - stable rotating spiral; 
OurBr - far-field break-up.}
\label{c3-phase2d}
\end{figure}
\begin{figure}
\includegraphics[width=0.45\textwidth]{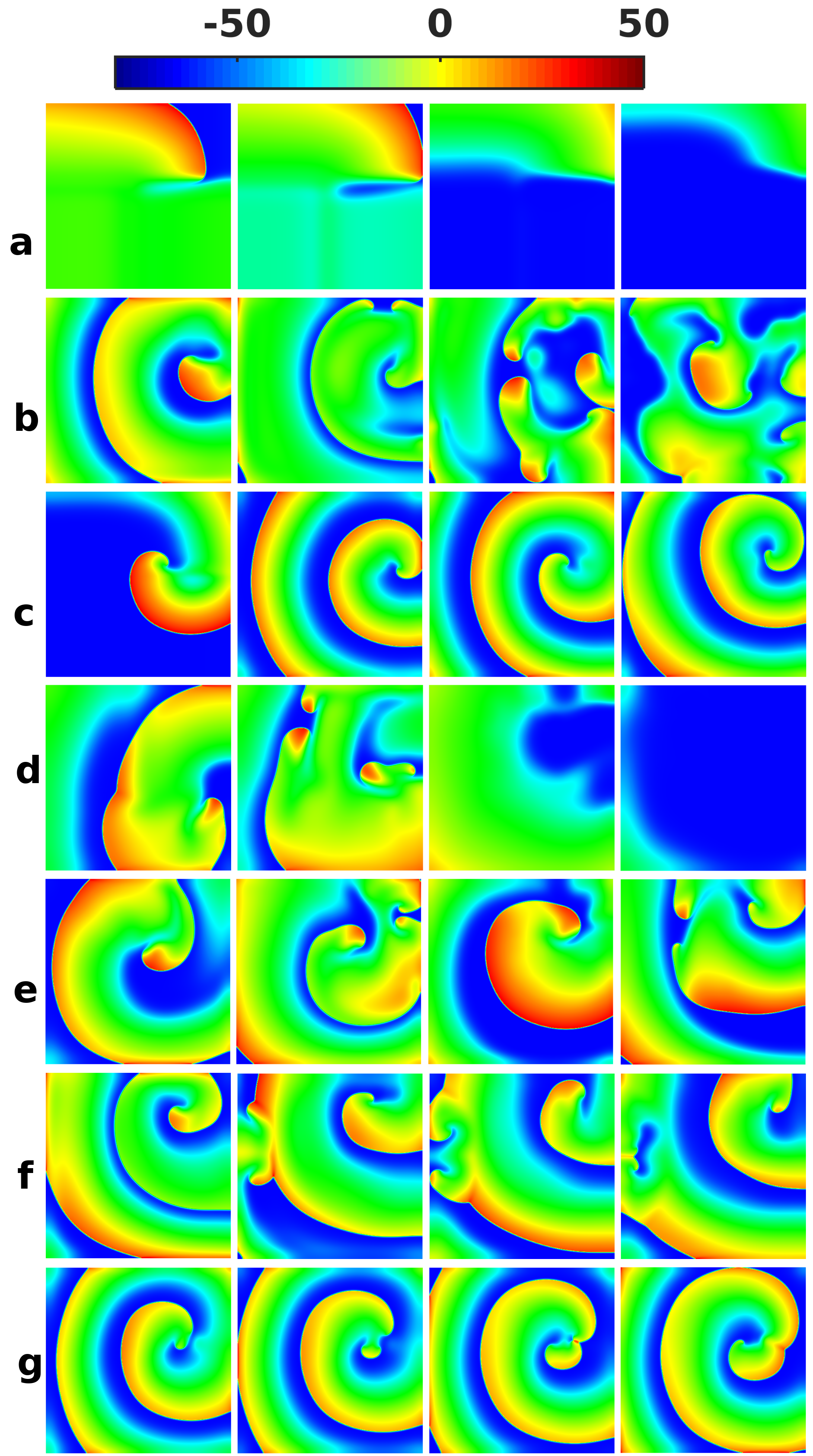}
\caption[Different kinds of wave dynamics in our DNSs for the 2D-HRD model, shown as pseudocolor plots of $V$]{
Pseudocolor plots of the transmembrane potential $V$ showing the
different kinds of wave dynamics in our DNSs for the 2D-HRD model. 
The sub-panels, from the left to the right, show the evolution of the 
system at four different points of time (for the complete spatiotemporal evolution see 
the Videos S35, S36, S37, S38, S39, S310 and S311 in the Supplementary Material~\cite{supplementary}).
We observe the following seven types of spiral-wave dynamics 
(seven panels from the top to the bottom):
(a) no spiral is formed ($N_{GKr}=2$, $N_{GCao}=10$); the waves disappear
before evolving into a spiral (magenta square $\mdblksquare$ in
Fig.~\ref{c3-phase2d}); (b) a stable state with broken spirals ($N_{GKr}=2$, $N_{GCao}=6$), 
($\mdblkdiamond$ in Fig.~\ref{c3-phase2d}); the broken spirals interacts and regenerate themselves, 
without disappearing; (c) a stable single rotating spiral ($N_{GKr}=7$, $N_{GCao}=7$), 
(green bubbles $\mdblkcircle$ in Fig.~\ref{c3-phase2d}); (d) an unstable break-up in
which the broken waves quickly disappear($N_{GKr}=1, N_{GCao}=1$), (red stars $\star$ in Fig.~\ref{c3-phase2d}); 
(e) region where a spiral breaks, recombines, rotates, again breaks and so on (green diamond, ${\mdblkdiamond}$, in Fig.~\ref{c3-phase2d}). 
(f) far-field break-up (observed for 
$N_{GKr}=5$ and $N_{GCao}=10$) [black-faced hexagon \ding{86} in Fig.~\ref{c3-phase2d}];
(g) a spiral with a local, short-lived instability of its core (here, $N_{GKr}=5$ and $N_{GCao}=9$), in the stable-rotating region
[in Fig.~\ref{c3-phase2d} the cyan colored bubbles ${\mdblkcircle}$].}
\label{c3-2dspiraldynamics}
\end{figure}

Given the range of parameter values we use, the initial spiral state shows
rich and varied spatiotemporal evolution. Here, we give a detailed description
of the various kinds of spiral-wave evolutions that we observe in the large
parameter space we investigate. 

We observe two (main) kinds of spiral arms, which we show in
Fig.~\ref{c3-spiral_arm}: (1) The first is a spiral with nonuniform arm widths at
different regions in the simulation domain; these regions can have inherent
instabilities, which cause spiral-arm thinning (but not enough to lead to the
breaking up of the spiral-arm). They are stable and preserve their shape. The
(average) wavelength of such a spiral wave is remarkably different from that of
the plane wave from which it is formed. This kind of spiral arm is formed
from plane waves of large width (as we find for low values of $G_{Kr}$ and
high values of $\gamma_{Cao}$). (2) A spiral with almost uniform arm width. 
These spirals, which are stable and retain their uniform arm-width, 
are formed from small-width plane waves (as we find for high values of $G_{Kr}$ and 
low values of $\gamma_{Cao}$). Here, the spiral wavelength is comparable to 
that of the plane wave. 

In summary, the spirals that form, in the different parameter regimes in the
HRD model, have different sizes and shapes, even though they might evolve in
a roughly similar manner to the final state and be characterized by generic
terms such as spiral break up or spiral rotation. In Fig.~\ref{c3-phase2d} we give
a phase diagram (or stability diagram) in the $N_{GKr}-N_{GCao}$ plane. [Recall
that $G_{Kr} = GKR \times N_{GKr}$, where $N_{GKr}= 1, 2, ...10$, and
$\gamma_{Cao} = GCAO/N_{GCao}$, where $N_{GCao}=1, 2...10$.] For this diagram we
have used the results from the $100$ different pairs of parameter values that
we have studied in the 2D HRD model. 
lot. The $3$ main
types of dynamics we can observe are:  broken waves, combined broken and rotating waves, 
and rotating waves. In each region we see some parts that sustain
stable waves, i.e., the waves stay in the medium; in some other parts there
are unstable waves, i.e., they move away and disappear from the medium. They
are shown with different markers in each colored region (details are given in the
figure caption). 

In the break-up region we see two different kinds of phenomena: waves breaking
up from the spiral core (core break-up); or waves breaking from the outer arms
(far-field break-up). The far-field break-up is weak because it does not spread
all the way to the core.

In Fig.~\ref{c3-2dspiraldynamics} we show pseudocolor plots of $V$, from
different stages of the spiral-wave evolution, going from left to right, for
each of the different kinds of wave dynamics. The time-evolution of the spiral
waves in the parameter-space we study is of the following seven major types
(shown in the seven rows of Fig.~\ref{c3-2dspiraldynamics}): (a) No spiral wave
is formed ($N_{GKr}=2$, $N_{GCao}=10$); the waves disappear from the medium
before evolving into a spiral [magenta square $\mdblksquare$ in
Fig.~\ref{c3-phase2d}].  (b) A stable state with broken spirals is obtained
($N_{GKr}=2$, $N_{GCao}=6$) [$\mdblkdiamond$ in Fig.~\ref{c3-phase2d}]; the
broken spirals interact and regenerate themselves, without disappearing. (c) A
stable single rotating spiral is formed ($N_{GKr}=7$, $N_{GCao}=7$) [green
bubbles $\mdblkcircle$ in Fig.~\ref{c3-phase2d}]. (d) There is unstable
break-up in which the broken waves quickly disappear (e.g., for $N_{GKr}=1,
N_{GCao}=1$) [red stars $\star$ in Fig.~\ref{c3-phase2d}].(e) A region where a
spiral breaks, recombines, rotates, again breaks and so on (green diamond,
${\mdblkdiamond}$, in Fig.~\ref{c3-phase2d}).  (f) There is far-field break-up
(observed for $N_{GKr}=5$ and $N_{GCao}=10$) [black-faced hexagon \ding{86} in
Fig.~\ref{c3-phase2d}]. (g) There is an instability in the core region,
exhibited by some parameter-combinations in the stable-rotating region (here,
$N_{GKr}=5$ and $N_{GCao}=10$) [in Fig.~\ref{c3-phase2d} the regions marked by
cyan colored bubbles ${\mdblkcircle}$]; such a core-break-up does not last, for
the spiral core quickly regenerates itself; this has no far-reaching effect on
the evolution of the spiral wave.

%
%

As we have mentioned earlier, the parameter region that we investigate in the
HRD model leads to spiral-wave dynamics that does not obey the restitution
hypothesis. According to this restitution hypothesis, if the slope of the APDR
curves is $ > 1$, the spirals break up; if the slope is $<1$, the spirals do not
break up. In our studies, the APDR slopes are always $ < 0.25$, yet we see
spiral breakup. However, we see a correlation between the maximal slopes of
the APDR restitution curves and the spiral-wave dynamics. The region where we
obtain stable rotating spiral wave dynamics corresponds to the lower-right
region of the APDR curves~\cite{supplementary}; this is where the slopes are
smallest, for the APDR curve is flat. The
APD gets smaller and smaller in this region. Small values of the APD corresponds to small
wavelengths $\lambda$. Such waves, with small widths, are not very prone
to break-up in the HRD model. 

\subsubsection{Dominant frequencies}

In this Subsection we examine the dominant frequencies of the spiral waves
that are formed in our 2D simulations of the HRD model. For this we use the
time series of the transmembrane potential $V$ from a few different sites in
the simulation domain. The power spectra of these time series yield the
dominant frequencies.
\begin{figure}
		\includegraphics[width=0.9\linewidth]{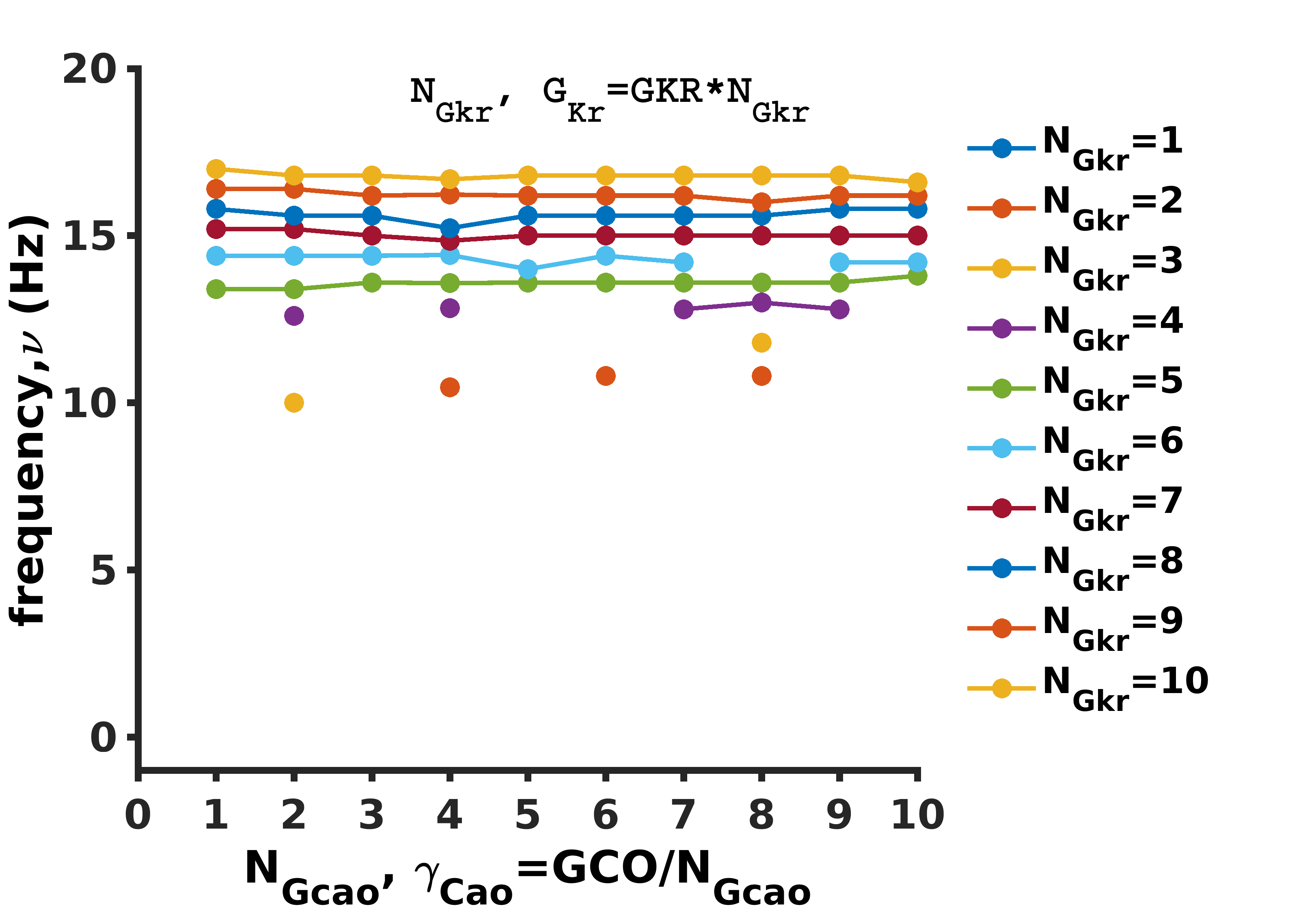}
\caption[The dominant frequencies that we obtain from power spectra for all our parameter values]
{The dominant frequencies that we obtain from power 
spectra  for all our parameter values, 
i.e., for $N_{GKr}=1, 2..10$ and $N_{GCao}=1, 2..10$. In the region where 
stable rotating waves exist, we see a prominent frequency in the spectrum. If the 
waves are unstable and disappear, it is not possible to identify a major frequency;
the missing parameter values in this plot correspond to these unstable regions.}
	\label{c3-twodpeakf}
\end{figure}
In Fig.~\ref{c3-twodpeakf} we show the dominant frequencies that we obtain from
power spectra for all our parameter
values, i.e., for $N_{GKr}=1, 2..10$ and $N_{GCao}=1, 2..10$. ({ We give the spectra corresponding to these different
wave dynamics in the Supplementary Material~\cite{supplementary}.})
In the region where
stable rotating waves exist, we see a prominent frequency in the spectrum. If
the waves are unstable and disappear, it is not possible to identify a major
frequency; the missing parameter values in this plot correspond to these
unstable regions. Note that the frequency of the rotating wave increases
as we increase $G_{Kr}$. The variation of $\gamma_{Cao}$ does not affect the
dominant frequency substantially, as is evident from Fig.~\ref{c3-twodpeakf}, where
the curves are almost flat.
\subsection{Three-dimensional (3D) Results}
\begin{figure}
		\includegraphics[width=0.9\linewidth]{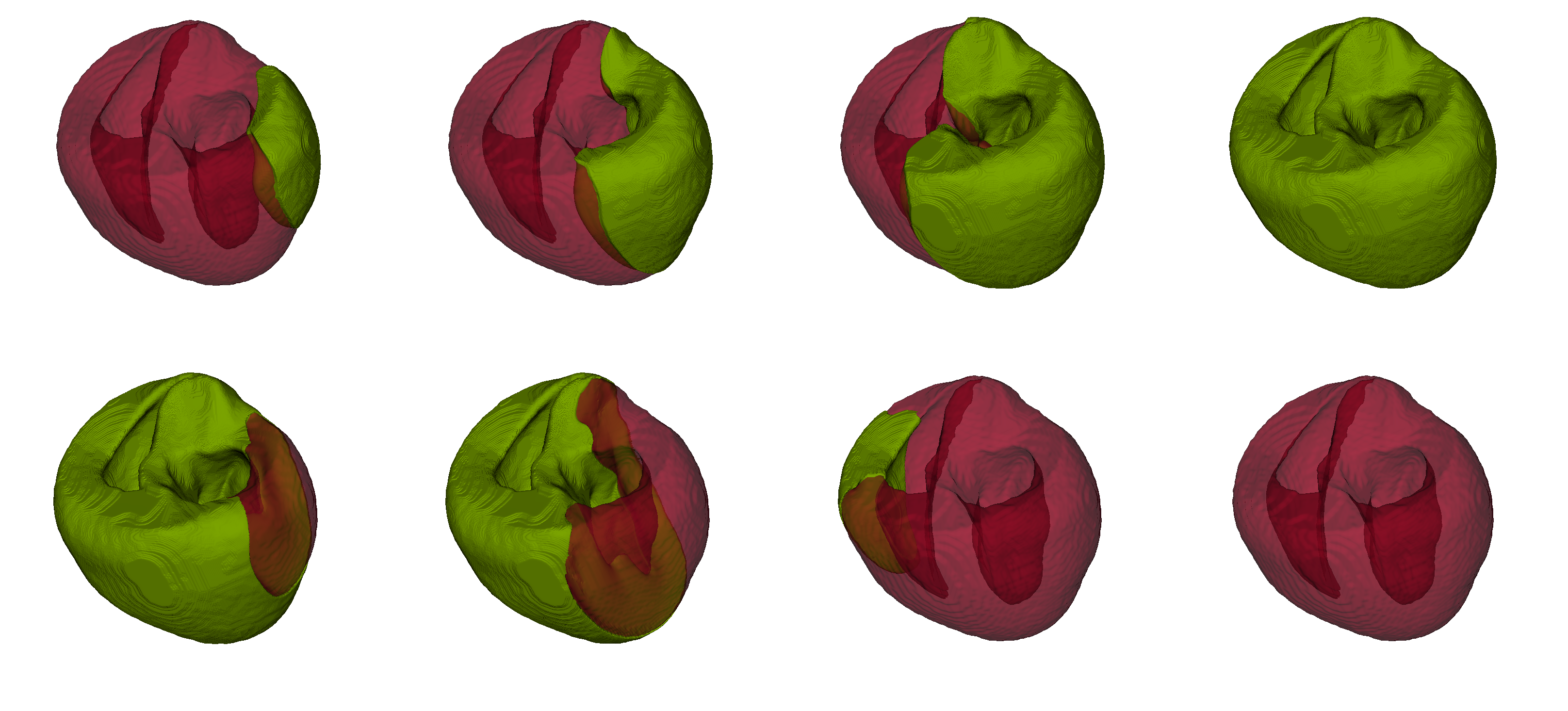}
\caption[A wave passing from one end to the other of our anatomically realistic
simulation domain geometry under normal conditions and without any
obstacles]{Two-level isosurface plots of $V$ for the HRD model illustrating a wave
passing from one end to the other of our anatomically realistic
simulation domain geometry under normal conditions and without any
obstacles. The upper panels show the initial stages, time increasing
from left to right; and the lower panels show the wave finally
disappearing from the domain. For the complete spatiotemporal
evolution see the Video S12 in the Supplementary Material~\cite{supplementary}.}
	\label{c3-3dplanewave}
\end{figure}
\begin{figure}
	\includegraphics[width=0.9\linewidth]{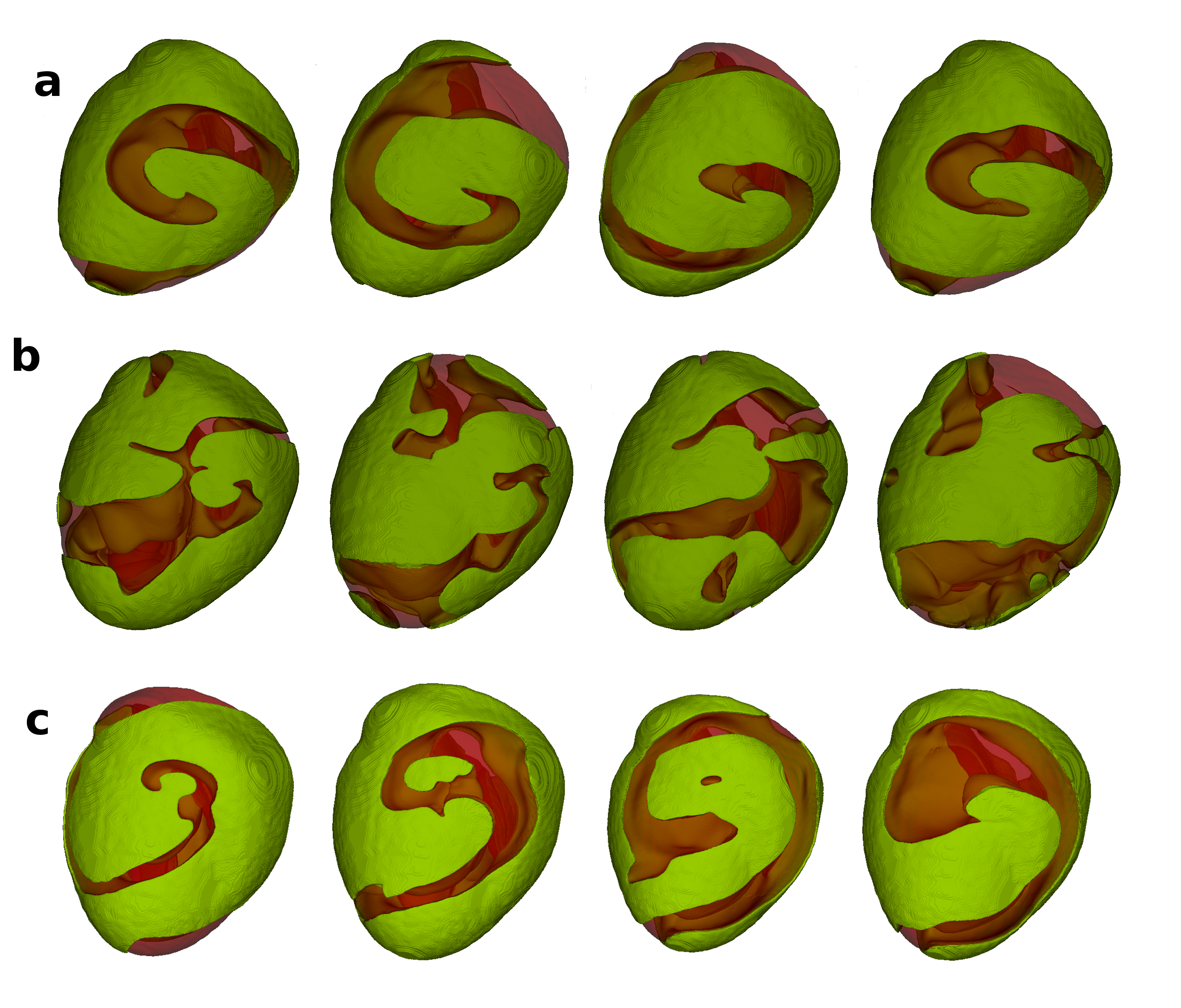}
\caption[Different types of scroll-wave dynamics in the 3D anatomical geometry of dog heart for our DNSs for the HRD model]
{Two-level isosurface plots of $V$ illustrating the following: (a) A scroll wave in the 
stable rotating state; it rotates with a slightly meandering core, without breaking up
($N_{GKr}=7; N_{GCao}=7$). (b) A meandering scroll wave breaks up and spreads 
through the domain leading to a statistically stable chaotic state, for the 
representative parameter set $N_{GKr}=1, \, N_{GCao}=1$.
(c) A scroll wave breaking while its central region passes through the right ventricle;
this broken part quickly recombines and continues as a rotating stable wave, thus 
showing that such anatomical structures have only short-term effects on the 
wave dynamics (here, $N_{GKr}=7;N_{GCao}=3$).
For the complete spatiotemporal evolution, see Videos S13, S14 and S15 in the Supplementary Material~\cite{supplementary}.}
\label{c3-3ddynamics}
\end{figure}
We now present our results on scroll-wave dynamics from our DNSs of the HRD
model on the realistic canine-heart geometry that we have described above. 

In Fig.~\ref{c3-3dplanewave} we show different stages of the spatiotemporal 
evolution of a plane wave passing through our anatomically realistic simulation
domain, from one end to the other end (here there are no obstacles). The lower panels
show how the wave moves and finally disappears. After a plane wave has passed through
this domain, we apply the S1-S2 cross-field stimulus to produce a scroll filament in 
the middle of the domain~\cite{supplementary}. We then vary the parameters $G_{Kr}$ and
$\gamma_{Cao}$ simultaneously and examine the effect of this change on
development of the scroll wave for $3$s in real time. 

With the original parameters of the HRD model, we see that the scroll wave
immediately breaks up and develops into a spatiotemporally chaotic state.
Within the time duration of our DNS, these broken scroll waves continue to
spread in the domain, interact, recombine, and break up without disappearing;
so the broken-wave state is statistically steady. The dynamics of broken
scrolls for the parameters $G_{Kr}=GKR$ and $\gamma_{Cao}=GCAO$ is shown in
Fig.~\ref{c3-3ddynamics}(b). If we decrease $\gamma_{Cao}=GCAO/N_{GCao}, $ by
using $N_{GCao}=3, 5, 7$, the spatiotemporal evolution of scroll waves is
qualitatively similar, with wave breaks and spatiotemporally chaotic behavior;
but, for $N_{GCao}=3$ and $7$, the waves are unstable, they meander, break up, 
and finally disappear. This behavior is not visible for $N_{GCao}=1$ and $5$.
Now we increase $G_{Kr}$ by factors of $3, \, 5$, and $7$. In each case we vary
$\gamma_{Cao}$ as above. Thus, we examine scroll-wave dynamics for $4\times4=16$ 
parameter sets. 
\begin{figure}
	\includegraphics[width=0.9\linewidth]{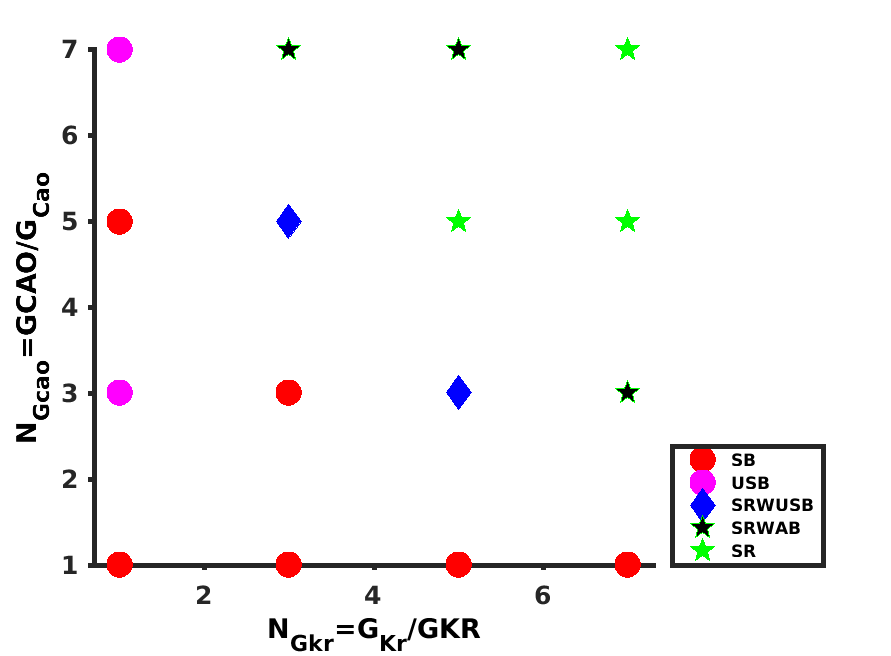}
	\caption[Phase diagram (or stability diagram) of scroll-wave dynamics in
the HRD model, in the 3D anatomically realistic geometry for a canine heart, and
in the $N_{GKr} - N_{GCao}$ plane]{Phase diagram (or stability diagram) of scroll-wave dynamics in 
the HRD model, in the 3D anatomically realistic geometry for a canine heart, and
in the $N_{GKr} - N_{GCao}$ plane, with $G_{Kr}=GKR\times N_{GKr}$ and 
$\gamma_{Cao}=GCAO/N_{GCao}$. We use markers:
a circle ($\mdblkcircle$) indicates scroll-wave break up; 
a star ($\medblackstar$) indicates a rotating scroll wave; and a diamond
($\mdblkdiamond$) indicates a rotating scroll with transient break-ups, 
in between (SRWUSB - Stable Rotation With Unstable Breakup).
The red circles indicate stable break-up (SB) and the magenta ones unstable break-up (USB).
In the rotating-scroll region, the green star indicates stable rotating scrolls (SR), 
the black star indicates a short-term anatomical break-up, which quickly recombines and 
continues as a rotating stable wave (SRWAB - Stable Rotation With Anatomical Breakup)
as shown in Fig.~\ref{c3-3ddynamics}(c).}
\label{c3-3dphase}
\end{figure}

The different kinds of scroll-wave developments that we observe in our DNSs of
the 3D HRD model in the anatomically realistic domain are depicted in
Fig.~\ref{c3-3ddynamics}: In Fig.~\ref{c3-3ddynamics}(a) we show the development of a
rotating scroll wave, with meandering, but without break up for a
representative parameter set ($N_{GKr}=7$ and $N_{GCao}= 7$). In
Figs.~\ref{c3-3ddynamics}(b) and (c) we show, respectively, how a meandering
scroll wave breaks up and spreads through the domain and how a scroll wave
breaks while its central region passes through the right ventricle, but such a
break-up goes away very soon, in a time $\simeq 100 ms$, and the wave recombines.
We see this anatomical
break-up arising solely because of the geometry rather than from functional
break-up associated with parameter dependence. This anatomical break-up
duration is negligible; it does not have a significant impact on the long-term
behavior of scroll-wave dynamics. 

The regions of stability of these different scroll-wave behaviors are depicted
in a phase diagram in Fig.~\ref{c3-3dphase} (see the figure caption for details).
Observe that, as we move to the right and upper regions of this phase diagram, 
scroll waves tend to be stable and rotating; they do not break up to form a
chaotic state. This region corresponds to $N_{GKr}=7, 5$ and $N_{GCao}=3, 5, 7$, 
and also $N_{GKr}=3$ and $N_{GCao} =5, 7$. And anatomical breakup of scroll
waves plays no significant role in the long-term dynamics of the system. Here
also, as in our study of the 2D HRD model, we obtain stable rotating scroll
waves in the bottom-right region of the APDR curves~\cite{supplementary}, where 
the APDR curve is nearly flat. 
\subsubsection{Dominant frequencies}

We now examine, as we did in 2D, the dominant frequencies of the scroll waves
that form in our DNSs of the 3D HRD model in the anatomical heart geometry.
We use time series of the transmembrane potential $V$ from a few different
sites from the simulation domain. 
{In the Supplementary Material~\cite{supplementary} we present
two-level isosurface plots of $V$ (left panel) showing different examples of scroll-wave
dynamics and the corresponding power spectra (middle panel) of the time series
(right panel) of $V$ from a representative point in the domain.} 
\begin{figure}
		\includegraphics[width=1\linewidth]{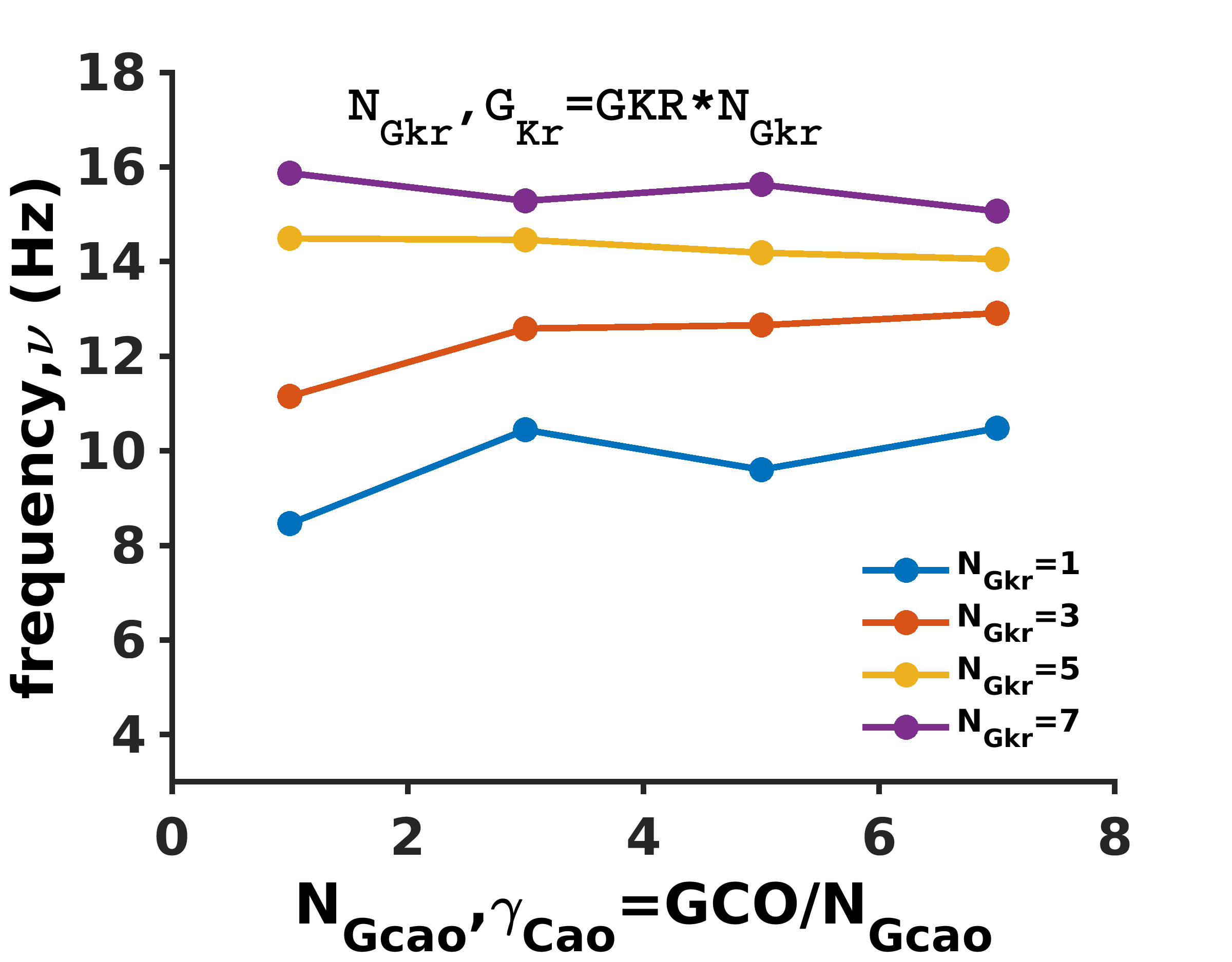}
		\caption[The dominant frequencies that we obtain from power
spectra for all our parameter values, $N_{GKr}=1, 2..7$ and $N_{GCao}=1, 2..7$]
{The dominant frequencies that we obtain from power 
spectra for all our parameter values, 
i.e., for $N_{GKr}=1, 2..7$ and $N_{GCao}=1, 2..7$. In the region where 
stable rotating waves exist, we see a prominent frequency in the spectrum. If the 
waves are unstable and disappear, it is not possible to identify a major frequency.
}
	\label{c3-threedpeakf}
\end{figure}

In Fig.~\ref{c3-threedpeakf} we plot the dominant frequencies that we obtain from
power spectra ({see the Supplementary Material~\cite{supplementary}}),
for $N_{GKr}=1, 2..7$
and $N_{GCao}=1, 2..7$. In the region where a stable rotating scroll wave
exists, we see a prominent frequency in the spectrum. If the waves are unstable
and disappear, it is not possible to identify a major frequency. 
Here also, 
as in the 2D case, the dominant frequency increases as $G_{Kr}$ increases; and
it is not affected significantly by a variation of $\gamma_{Cao}$. 

\section{Human-Heart (TP06) Model}
\label{c3-sec:HHResults}
We repeat the whole procedure described above on a human-heart geometry, with fiber
orientation, by using the TP06 human-ventricular model. Our goal is to compare
these results with their HRD-model counterparts. We present our
results in {two subsections, from 2D and 3D simulations.}
{The shapes of the action potential (AP) and the 
APD-restitution (APDR) curves, for $5$ representative cases for the 
rapid rectifier current $I_{Kr}$, the  Calcium current $I_{CaL}$,and 
different values of $G_{CaL}$ and $G_{Kr}$, for the TP06 model, are given in the 
Supplementary Material~\cite{supplementary}.}
\begin{figure}
\includegraphics[width=1\linewidth]{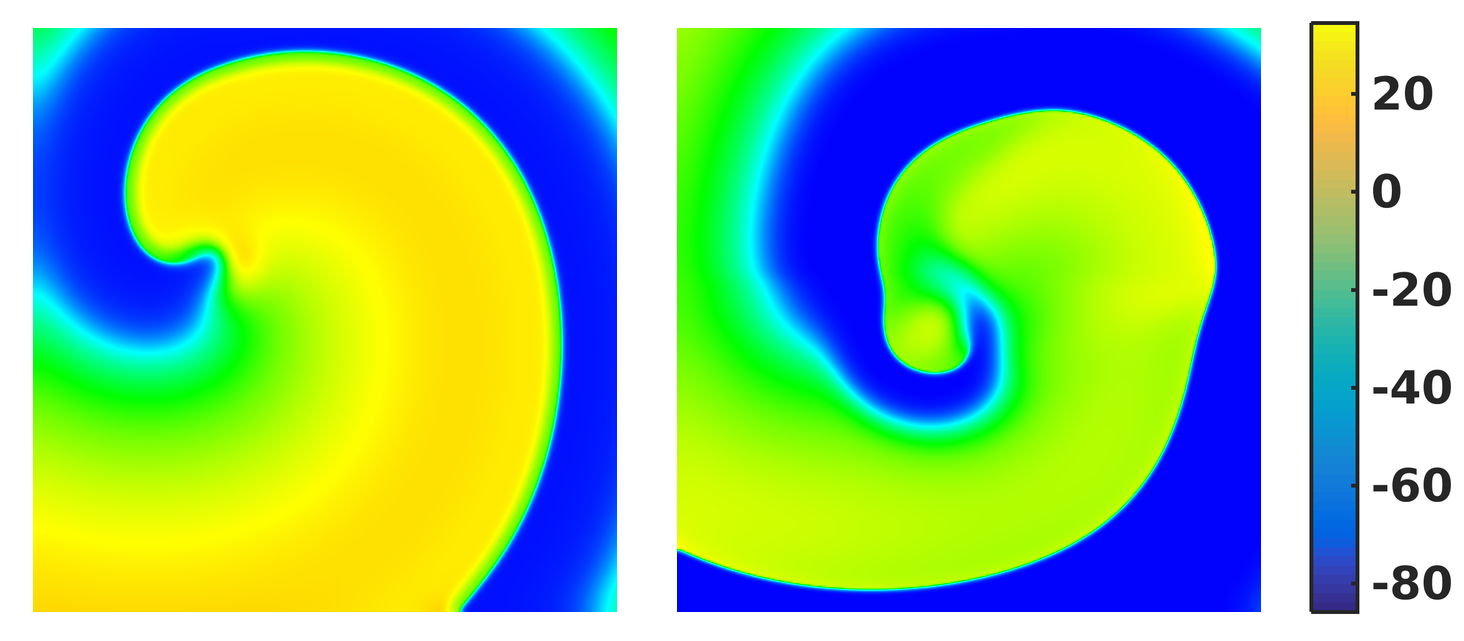}
\caption[Two major types of spiral arms that we observe in our 2D TP06-model simulations]
{Pseudocolor plots of the transmembrane potential $V$ showing the two major 
types of spiral arms we observe in our 2D TP06-model simulations. 
Left panel: a spiral with an almost uniform arm width. Right panel: 
a spiral with a nonuniform wave width in different regions, with thin arms
in some regions; (cf. Fig.~\ref{c3-spiral_arm} for the HRD model).
For the complete spatiotemporal evolution see the Videos S16, S17 in the Supplementary Material~\cite{supplementary}. } 
\label{c3-hspiral_arm}
\end{figure}
\subsection{2D Results}

For our simulations of the 2D TP06 model, we follow the same methods, 
numerical schemes, and initial spiral conditions we have used in our
HRD-model DNSs. The conduction velocity of a plane wave passing from one end
to the other end in the 2D domain is the same for all the cases, which is
$6.960 m/s$ for the TP06 model. 
We have measured the wavelength of the plane
As we have
described in Sec.\ref{c3-sec:methods}, we vary $G_{Kr}=GKR \times N_{GKr}$ and
$G_{CaL}=GCaL/N_{GCaL}$ to obtain a total of $20$ different cases.
\begin{figure}
\includegraphics[width=1\linewidth]{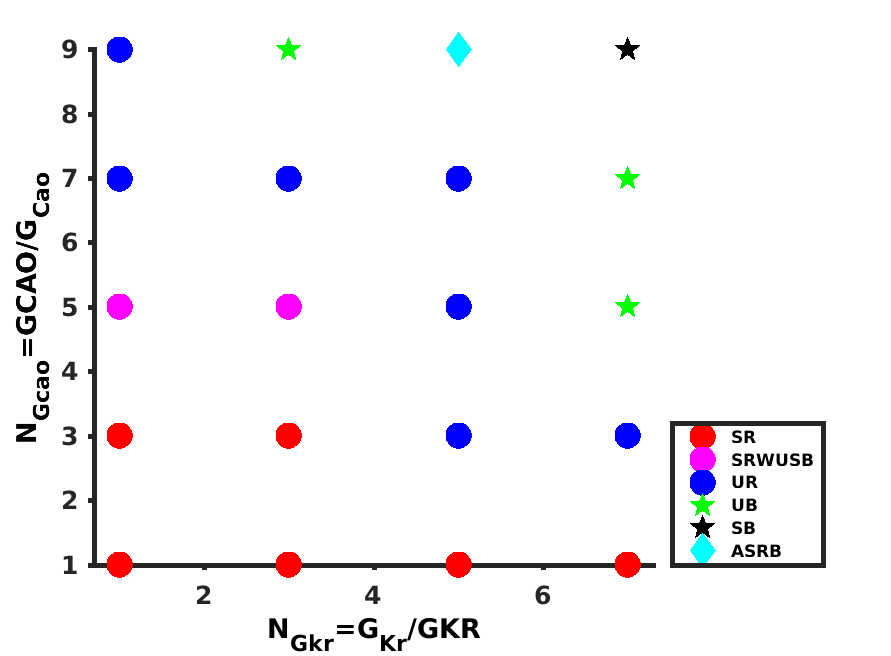}

\caption[Phase diagram (or stability diagram) in the $N_{GKr}-N_{GCaL}$ plane
for the types of spiral-wave dynamics in the 2D TP06 model, for the region of parameter space in our study]
{Phase diagram (or stability diagram) in the $N_{GKr}-N_{GCaL}$ plane
for the types of spiral-wave dynamics in the 2D TP06 model, for the
region of parameter space in our study. There are regions of broken-spiral states, 
rotating-spiral states, unstable-rotating states, and a mixed state. Different markers
distinguish the following:  Stable rotating spirals are denoted by bubbles (${\mdblkcircle}$) with
different face colors; we see broken spirals
(\ding{72}),  and unstable
rotation (blue bubble). At one parameter value, denoted by a diamond
($\mdblkdiamond$), we see an alternating state, in which the wave
undergoes a transition from a spiral to a broken-wave state and then
recombines into a spiral again; this alternation continues for the
duration of our DNS. In each region we see (a) stable states, in which
the waves persist in the domain, and (b) unstable states, in which the
waves move away and disappear completely from the domain. We define the
following acronyms: SR - stable rotating; SRWUSB - stable rotating with
unstable break-up in between; ASRB - alternating stable-rotating and
breaking states; UR - unstable rotation; UB - unstable break-up; and SB
- stable break-up (cf. Fig.~\ref{c3-phase2d} for the 2D HRD model).}
\label{c3-hphase2d}
\end{figure}
\begin{figure}
\includegraphics[width=1\linewidth]{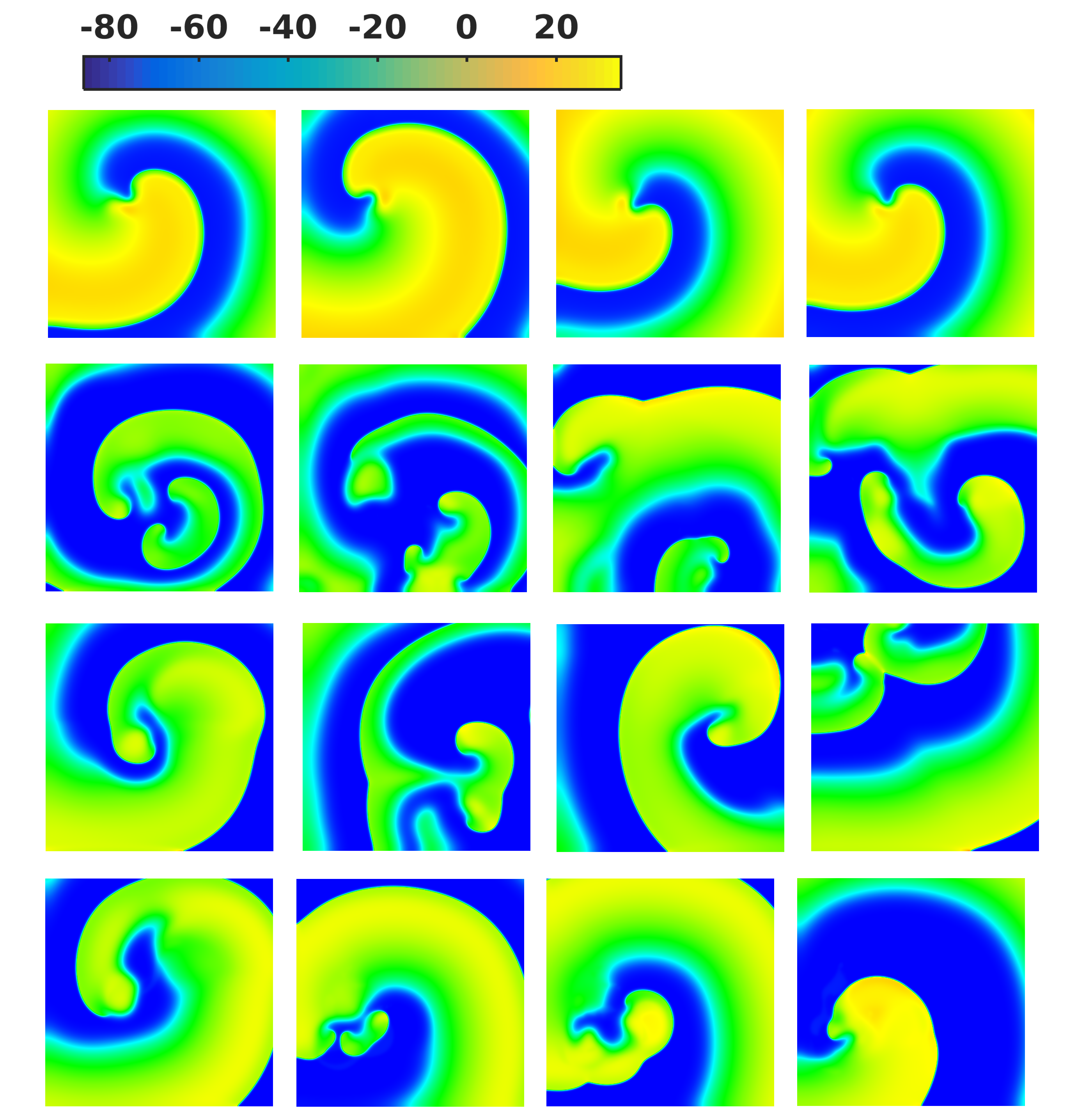}

\caption[The four different kinds of spiral-wave dynamics that we observe in our DNSs for the 2D TP06 model]
{Pseudocolor plots of $V$ illustrating the four different kinds of spiral-wave 
dynamics that we observe in our DNSs for the 2D TP06 model.
(for the complete spatiotemporal evolution see the Videos S18, S19, S20 and S21 in the Supplementary Material~\cite{supplementary}): 
(a) A stable, single rotating spiral state (here, $N_{GKr}=1$, $N_{GCaL}=3$);
red bubbles($\mdblkcircle$)in the phase diagram Fig.~\ref{c3-hphase2d}.
(b) A stable state with broken spirals (here, $N_{GKr}=7$, $N_{GCaL}=9$);
a black star ($\star$) in Fig.~\ref{c3-hphase2d}.
(c) Alternating rotating and broken-wave states (here, $N_{GKr}=5$ and $N_{GCaL}=9$);
in Fig.~\ref{c3-hphase2d} a cyan diamond,${\mdblkdiamond}$. 
(d) The central spiral core shows an instability and a tendency to break up; 
such break-ups are transients, for the center quickly regenerates itself; 
this has no far-reaching effect on the evolution of spiral wave and it remains as a 
localized, isolated event at the center; magenta bubbles (${\mdblkcircle}$) in 
Fig.~\ref{c3-hphase2d} (cf. Fig.~\ref{c3-2dspiraldynamics} for the 2D HRD model).
}
\label{c3-h2ddynamics}
\end{figure}

We now describe the types of spiral-wave dynamics that we obtain in the large
parameter space we have studied in the 2D TP06 model. As in the canine 2D
HRD model (Fig.~\ref{c3-spiral_arm} in Sec.\ref{c3-dog2d}), we observe two types of
spiral arms. We depict these in Fig.~\ref{c3-hspiral_arm}. 

In Fig.~\ref{c3-hphase2d} we give the phase diagram (or stability diagram) for the
different types of spiral-wave dynamics we obtain in the 2D TP06, for the
$20$ cases that we have studied. The $3$ main colors in this phase diagram
denote broken waves, a phase with both broken and rotating spirals, and
rotating waves. In each region, we find that some waves are stable (they
repeat themselves in time and do not move away from the domain), whereas others
are unstable (they move away and disappear from the domain); these waves are
shown with different markers in each coloured region. As we go from left to
right in this phase diagram, the meandering of the wave increases, so it is
prone to break up; eventually, for high values of $G_{Kr}$, the wave does break
up. [Figure~\ref{c3-hphase2d} is the counterpart of Fig.~\ref{c3-phase2d} for the
HRD model.]

In Fig.~\ref{c3-h2ddynamics} we show the different kinds of spiral-wave dynamics
that we observe in our DNSs for the 2D TP06 model. The panels from the left
to the right represent the evolution of the system at four different points of
time. We observe the following four types of spiral-wave dynamics. (a) A
stable, single rotating spiral state (for $N_{GKr}=1$, $N_{GCaL}=3$); stable
spiral states are represented by red bubbles ($\mdblkcircle$) in the phase
diagram of Fig.~\ref{c3-hphase2d}. (b) A stable state with broken spirals (here, 
$N_{GKr}=7$, $N_{GCaL}=9$); the broken spirals interact with each other, 
regenerates themselves, and do not leave the domain. This is denoted by a
black star ($\star$) in Fig.~\ref{c3-hphase2d}. (c) We see alternately rotating
and breaking spirals (here, for $N_{GKr}=5$ and $N_{GCaL}=9$); in
Fig.~\ref{c3-hphase2d}, a cyan diamond,${\mdblkdiamond}$. (d) In some parts of the
stable-rotating-wave region (green regime in Fig.~\ref{c3-hphase2d}), some of the
spiral cores show a tendency to break and recombine in a very short interval
of time. This instability of the core does not affect the long-term stability
and dynamics of the spiral; the spiral core quickly reassembles and continues
to rotate (see Fig.~\ref{c3-h2ddynamics}(d)), and remains as a localized, 
isolated event at the center; the regions in which we observe this are marked by
magenta bubbles (${\mdblkcircle}$) in Fig.~\ref{c3-hphase2d}. In addition to the
states mentioned above, we find unstable rotating states, for parameters
denoted by blue bubbles, and unstable broken spiral waves, for parameters
marked as green stars, in the phase diagram Fig.~\ref{c3-hphase2d}. These states
last only for a short duration and quickly move away from the domain.
[Figure~\ref{c3-h2ddynamics} is the counterpart of Fig.~\ref{c3-2dspiraldynamics} for
the HRD model.]
\begin{figure}
		\includegraphics[width=1\linewidth]{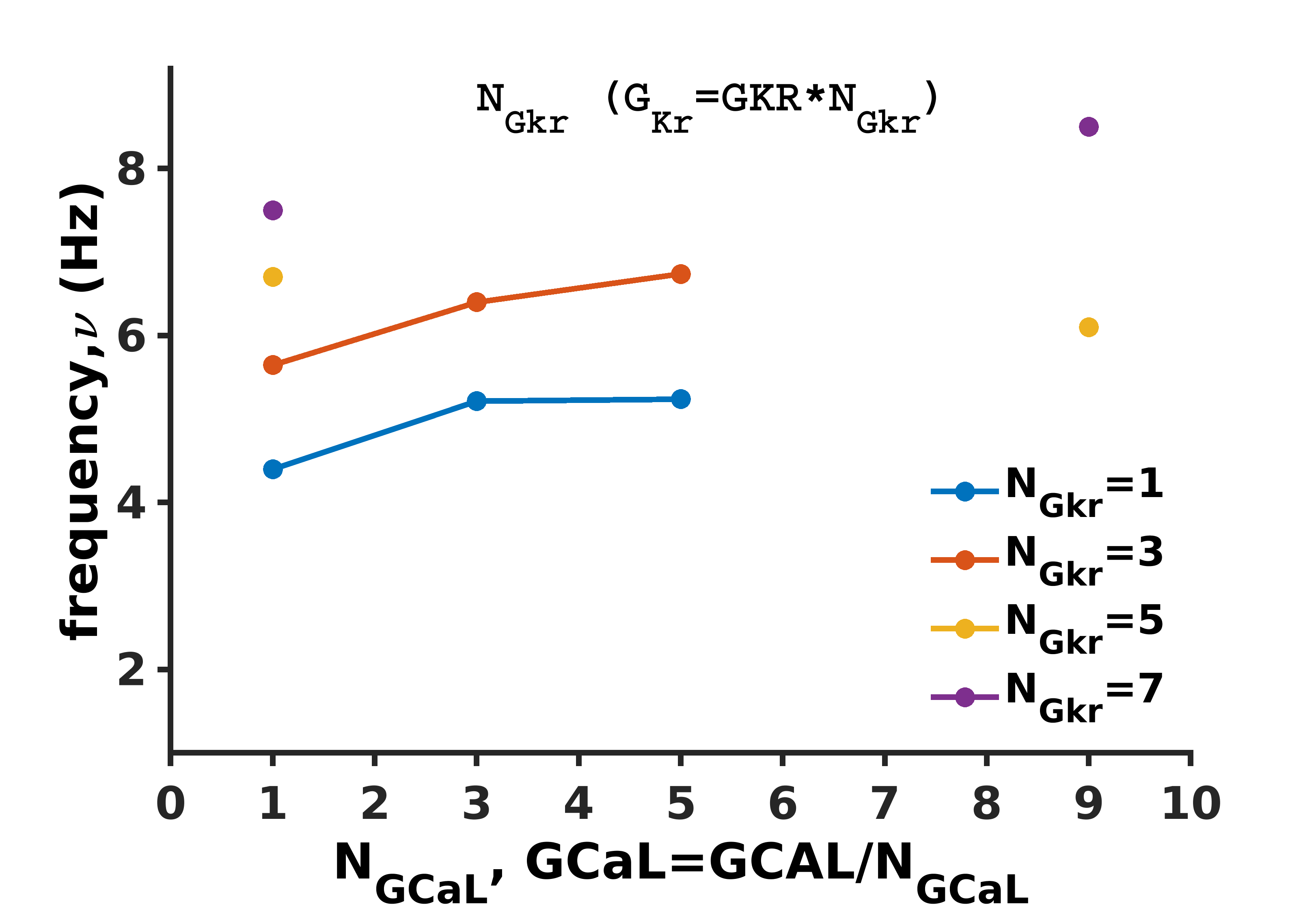}
\caption[The dominant frequencies that we obtain from power
spectra for all our parameter values; $N_{GKr}=1, 3..7$ and $N_{GCaL}=1, 3..9$]
{The dominant frequencies that we obtain from power 
spectra for all our parameter values
(plots versus $N_{GCaL}$ for all the parameter values; $N_{GKr}=1, 3..7$ and 
$N_{GCaL}=1, 3..9$). In the region where 
stable rotating waves exist, we see a prominent frequency in the spectrum. If the 
waves are unstable and disappear, it is not possible to identify a major frequency;
the missing parameter values in this plot correspond to these unstable regions (cf.
Fig.~\ref{c3-twodpeakf} for the 2D HRD model).}
\label{c3-htwodpeakf}
\end{figure}
\subsubsection{Dominant frequencies}

In this Subsection we examine the dominant frequencies of the spiral waves formed
in our 2D TP06 simulations. As we did in our study of the HRD model, we
identify these frequencies from the power spectra of the time series of the
transmembrane potential $V$, which we obtain from a few different points in the
simulation domain. 

In Fig.~\ref{c3-htwodpeakf} we portray the most dominant frequencies
that we observe for each of the parameter values. The figure is discontinuous
in the unstable regions because the unstable regions do not sustain a wave. In
the regions where the spiral has a well-defined peak frequency, the frequency
increases as we move to high values of $G_{Kr}$; by contrast, $G_{CaL}$ has a
very mild effect on the dominant frequency.
{A few examples of wave dynamics, along with
the corresponding power spectra and the time series of $V$, are shown in the 
Supplementary Material~\cite{supplementary}.}
\begin{figure}
 \includegraphics[width=0.9\linewidth]{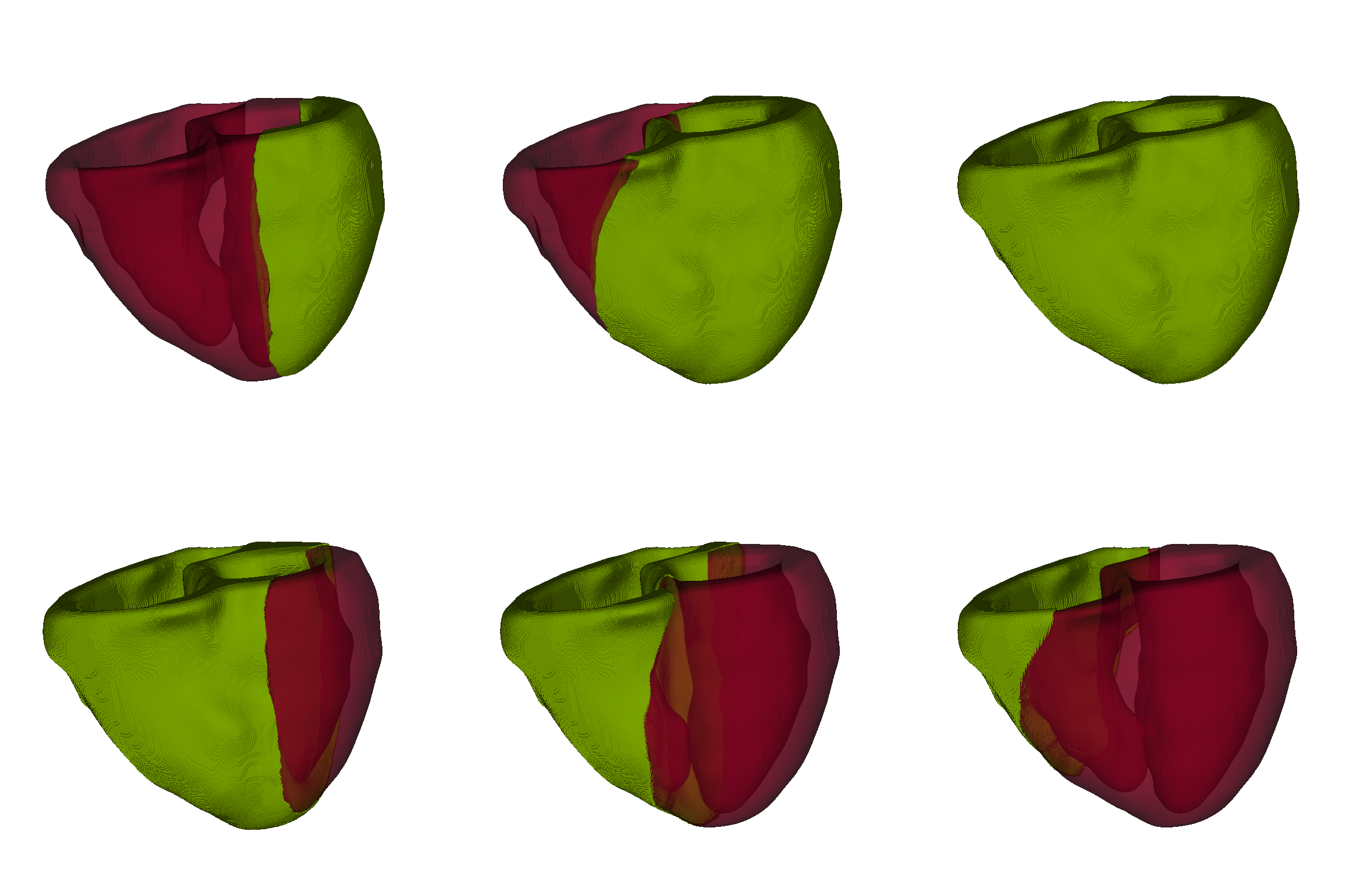}
	\caption[A wave passing from one end to the other of our 3D realistic, 
 human-heart geometry in the normal situation (without any obstacles), for the TP06 model.]
 {Two-level isosurface plots of $V$ for the 3D TP06 model illustrating
	a wave passing from one end to the other of our 3D realistic, 
	human-heart geometry in the normal situation (without any obstacles). 
	The upper panels show the
	initial stages from left to right; the lower panels show the wave
	finally moving out of the medium (cf. Fig.~\ref{c3-3dplanewave}
	for the 3D HRD model). For the complete spatiotemporal 
	evolution see the Videos S22 in the Supplementary Material~\cite{supplementary}. } 
\label{c3-h3dplanewave} 
\end{figure}
\subsection{3D Results}
We now present our results from our DNSs of the TP06 model in a realistic, 
human-heart geometry that is reconstructed from DTMRI data (see
Fig.~\ref{c3-hanat_geom}). 
\begin{figure}
	\includegraphics[width=1\linewidth]{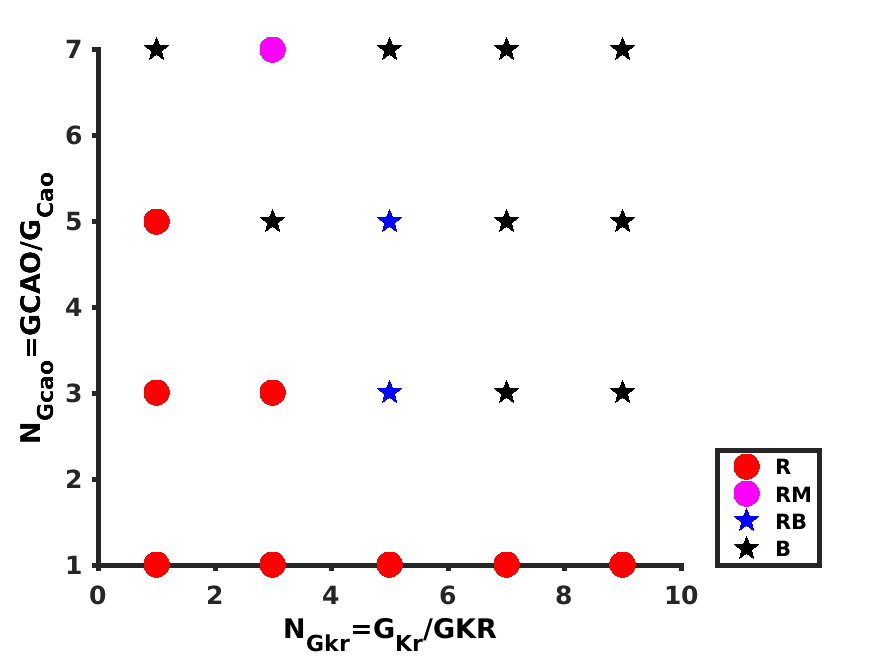}
	\caption[The phase diagram for scroll-wave dynamics in 3D TP06
 model in a realistic, human-heart geometry]
	{The phase diagram for scroll-wave dynamics in 3D TP06
	model in a realistic, human-heart
	geometry. This diagram uses  markers in the
	$N_{GKr}-N_{GCaL}$ parameter space. In general, a bubble indicates a
	rotating scroll wave and a star indicates scroll break-up. The red
	circles indicate stable rotation (R) and the magenta circle indicates
	a highly meandering rotating scroll (RM). The blue stars indicate a
	rotating scroll, which breaks up eventually (RB); and the black stars
	indicates stable break-up (B). All phases are stable states here, i.e., 
	the waves do not disappear from the medium (cf. Fig.~\ref{c3-3dphase}
	for the 3D HRD model).} 
\label{c3-h3dphase}
\end{figure}

In the normal situation, a plane wave passes from one end to the other end of
this geometry (without any obstacle). Figure~\ref{c3-h3dplanewave} shows different
stages of such a passing of a plane wave through the human-ventricular
geometry. The upper panels show the initial stages; and the lower panels show
the final stages when the wave finally moves away from the domain.
\begin{figure} \includegraphics[width=1\linewidth]{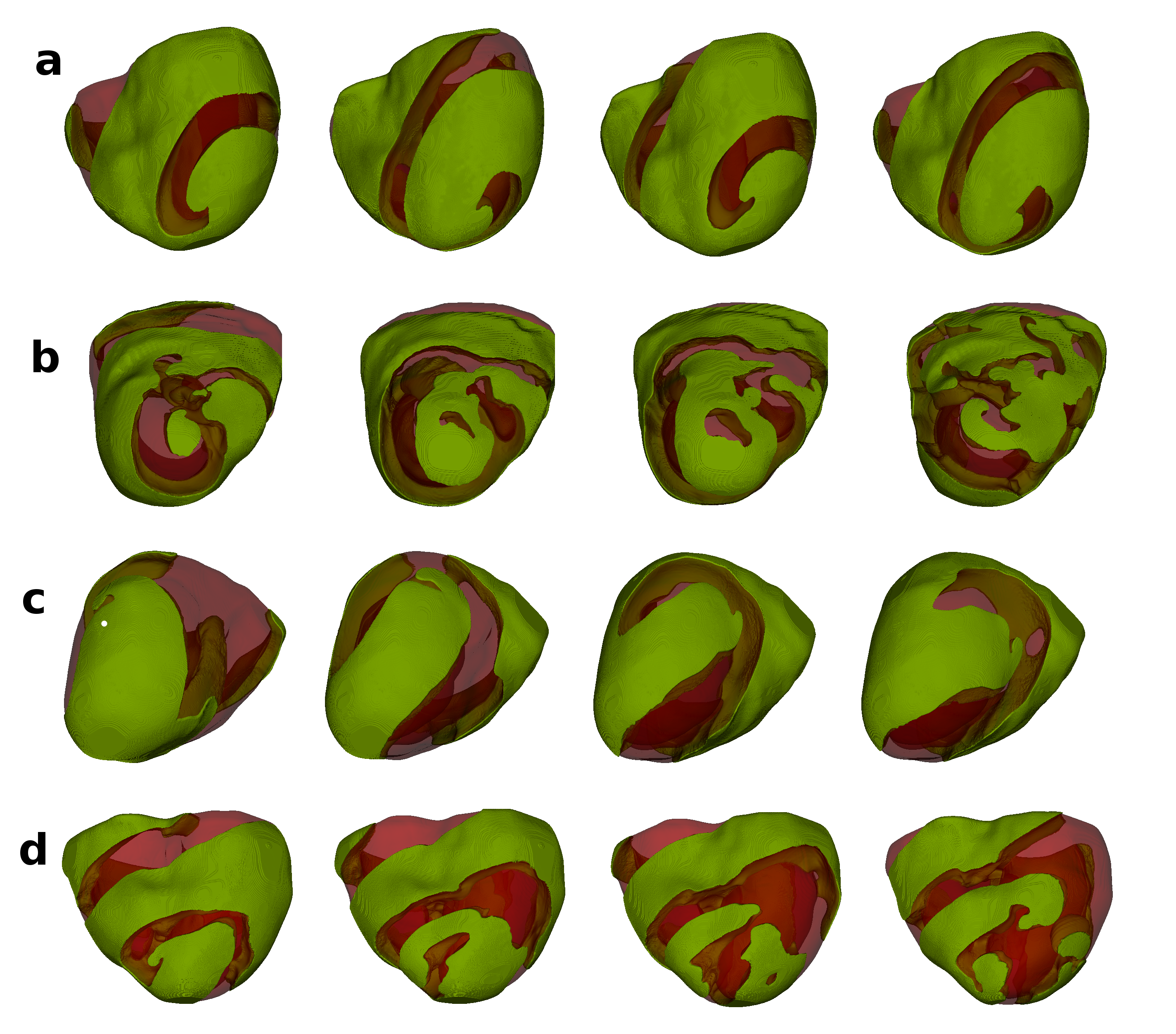} 
	\caption[The four types of scroll wave dynamics we observe in our 3D system for TP06 model]
	{Representative figures showing each of
	the four types of scroll wave dynamics we observe in our system. The
	panels from left to right show, via two-level isosurface plots of $V$, the 
	spatiotemporal evolution of the
	scrolls. (a) A stable rotating state where the scroll wave rotates and
	slightly meanders, without breaking up ($N_{GKr}=5, N_{GCaL}=1$). 
	(b) A stable chaotic state where a scroll wave breaks up and the broken
	scrolls spread throughout the medium ($N_{GKr}=7, N_{GCaL}=3$). 
	(c) A scroll wave meandering throughout the medium ($N_{GKr}=3, 
	N_{GCaL}=7$). (d) A scroll wave rotates for a long while and
	eventually breaks up ($N_{GKr}=3, N_{GCaL}=3$); (cf. Fig.~\ref{c3-3ddynamics}
	for the 3D HRD model). For the complete 
	spatiotemporal evolution see the Videos S23, S24, S25 and S26 in the Supplementary Material~\cite{supplementary}. } 
	\label{c3-h3ddynamics}
\end{figure}

We follow the same procedure that we have used above for the HRD canine-ventricular 
model to initiate a scroll wave. Then, as in the HRD case, we vary the parameters
$G_{Kr}$ and $G_{CaL}$ simultaneously and investigate the effects of these
changes on the scroll-wave development. We record this development for $2.5$
seconds in real time. 

With the original parameters of the TP06 model, we see that the scroll wave
rotates without breaking; and it is stable for a long time after its
initiation. The dynamics of such a scroll, for the parameters $G_{Kr}=GKR$ and
$G_{CaL}=GCAL$, is shown in Fig.~\ref{c3-h3ddynamics}a. 

With a decrease of $G_{CaL}$ by factors of $3, \, 5, $ and $7, $ the dynamics we
observe is the same as above; a stable rotating wave is formed and it continues
to rotate, with a small amount of meandering, but it does not break. We then
increase $G_{Kr}$ by factors of $3, \, 5, \, 7, $ and $9$, while simultaneously
varying $G_{CaL}$ as above. Thus, we examine $5\times4=20$ parameter sets. We
present the general results of the scroll-wave behaviors in the phase diagram
(or stability diagram) of Fig.~\ref{c3-h3dphase}. We observe that, as we move to
the right and upper region of this phase diagram, the scroll waves tend to
break up. This region corresponds to $N_{GKr}=7, \, 9$ and $N_{GCaL}=3, \, 5$
and $ 7$, and also $N_{GKr}=5, N_{GCaL}=7$ and $N_{GKr}=3, N_{GCaL} =5$.
[This is the counterpart of Fig.~\ref{c3-3dphase} for the 3D HRD model.]

Figure~\ref{c3-h3ddynamics} shows the scroll wave's time development for each type
of dynamics that we observe: (a) shows a stable rotating scroll-wave; and (b) depicts
a scroll wave breaking up and forming a chaotic state ($N_{GKr}=9, N_{GCaL}=
5$). We observe another interesting case in the scroll-wave
rotating-meandering region, for $G_{Kr}=3$ and $G_{CaL}=7$: the wave meanders
widely, throughout the simulation geometry; this is shown in
Fig. \ref{c3-h3ddynamics}(c); (d) provides an example of the cases $N_{GKr}=5
N_{GCaL} = 3, 5$, where the scroll-wave rotates for a long time, and finally
breaks up after $\approx 2$s. [This is the counterpart of
Fig.~\ref{c3-3ddynamics} for the 3D HRD model.]

By comparing these figures with their 3D-HRD-model counterparts, we can see
that scroll-wave dynamics for the TP06 model is quite different from that in
the HRD model (for corresponding parameter regions). 
\subsubsection{Dominant frequencies}

In this Subsection we examine the dominant frequencies of the scroll waves formed
in the anatomical heart geometry by using the time series of the transmembrane
potential from a few different sites of the domain. 
\begin{figure}
		\includegraphics[width=1\linewidth]{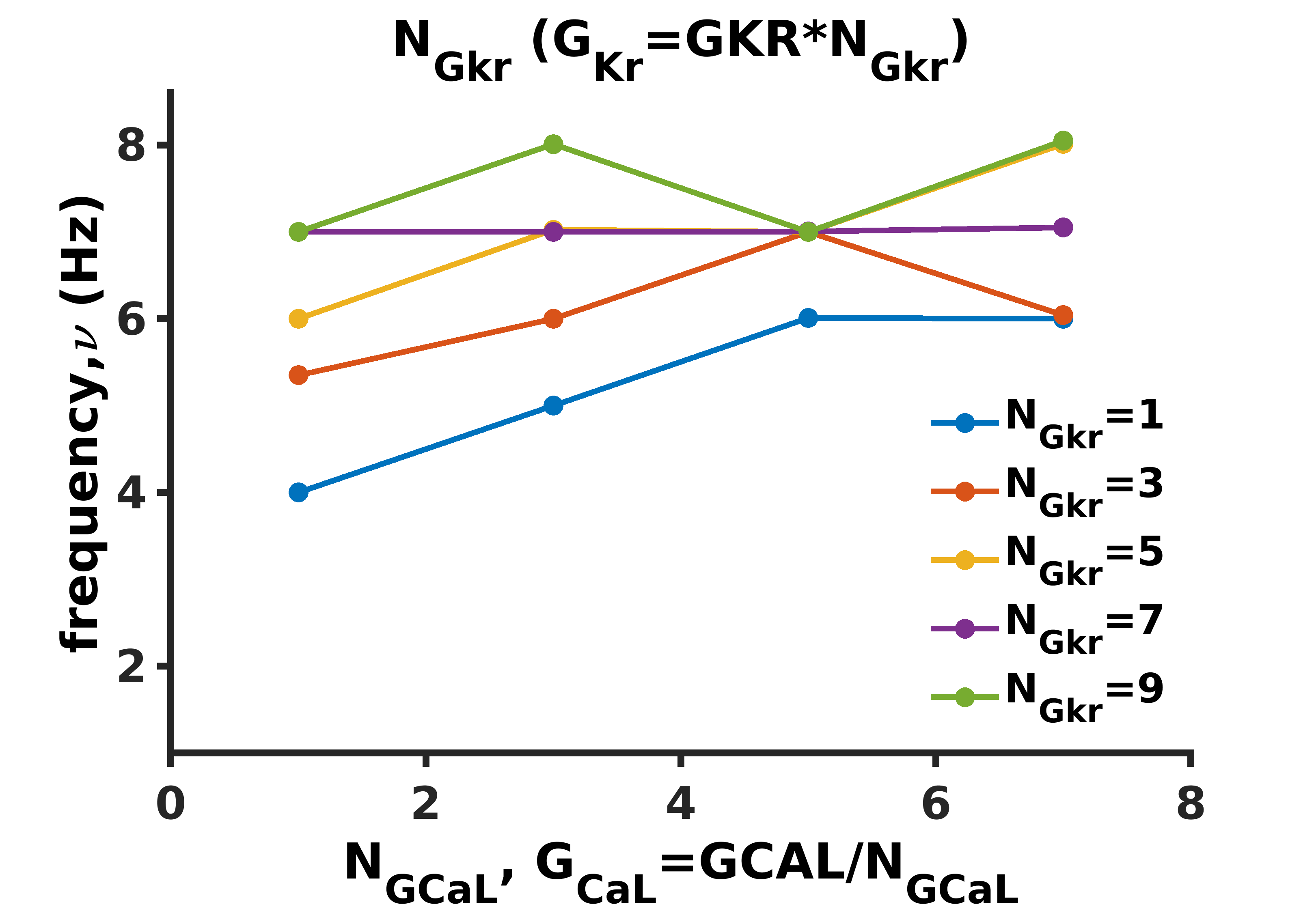}
\caption[The dominant frequencies we observe for all the parameter values, 
$N_{GKr}=1, 2..9$ and $N_{GCaL}=1, 2..7$, for TP06 model]
{The dominant frequencies we observe for all the parameter values;
$N_{GKr}=1, 2..9$ and $N_{GCaL}=1, 2..7$. For stable states, the
frequency increases with $G_{Kr}$; the variation with $G_{GaL}$ is
	either small or nonexistent. }

\label{c3-hthreedpeakf}
\end{figure}

Here also, as in the 2D case, the dominant frequency of the scroll wave
increases with an increase in $G_{Kr}$; but this frequency is not affected significantly
by the variation of $G_{CaL}$. This is shown in Fig.~\ref{c3-hthreedpeakf}, where
we show the dominant frequencies of the scrolls for the whole parameter space
that we explore.
{A few examples
of scroll-wave dynamics, along with the corresponding power spectra and the time
series of $V$ for the 3D TP06 model, are given in the Supplementary Material~\cite{supplementary}}.
\section{Conclusions}

We have conducted extensive, \textit{in silico} studies of the direct effects
of two major, ion-channel conductances on spiral- and scroll-wave dynamics in
idealised (2D) and anatomically detailed (3D) geometries for the canine- and
human-ventricular models (HRD and TP06, respectively). We find that
$I_{GKr}$ and $I_{GCaL}$ are two important currents that determine the
characters of spiral- and scroll-wave dynamics, namely, rotation, meandering, or
the break-up of these waves in the ventricles. The effects of changes in these
currents, on such wave dynamics, are most clearly visible when they are changed
together, rather than individually (specifically, when $G_{Kr}$ increases and
$G_{CaL}$ decreases). We have observed this qualitative feature in the two
distinctly different models, for two different mammalian species, namely the
canine-ventricular HRD model and human-ventricular TP06 model. 

The precise forms of the spiral or scroll wave that are formed in the
simulation domain, in the absence of any parameter variation, are, of course, 
model specific. Hence, changes in these waves and their dynamics, as a function
of model parameters, are also model specific. Nevertheless, there is a
qualitative similarity in the transitions from one phase to another, the phase
diagrams that we have presented for both HRD and TP06 models in both 2D and
3D. Details differ, of course, as we can see by comparing the phase diagrams
for these models carefully. Such a detailed comparison between wave dynamics in these 
different mammalian models has not been attempted hitherto.

In our HRD model simulations, the scroll-wave break-up that we observe, 
without varying the values of $G_{Kr}$ and $G_{Cao}$, undergoes a transition to
a stable meandering wave, without break up, as we vary the parameters. In the
TP06 model simulations, we observe the reverse phenomenon, i.e., a transition
occurs, from the stable rotating state, in the initial parameter region, to a
broken-scroll or chaotic state (even though we vary parameters over a range
that is similar to the one we use in our HRD-model simulations). In both
these models, the combination of the ion-channel conductances, for $I_{Kr}$ and
$I_{GCaL}$, plays a crucial role in determining the nature of scroll-wave
dynamics. Recall that the parameter region we have explored is devoid of other
commonly observed mechanisms that lead to uncontrolled scroll-wave behavior, 
such as a sharp APD restitution curve, early after depolarizations, and delayed
after depolarizations.

There is no consensus on whether the geometric details of the heart itself
affect scroll-wave dynamics or not; and, if they do, to what extent and in
which way. We have observed, in our simulations with anatomically realistic geometries
and fiber-orientation details, that the long-term effect of the geometry, 
without abnormal inhomogeneities or other variations, on scroll-wave dynamics 
is negligible. The role that the geometry itself plays here is to
trap the re-entrant waves, preventing them from decaying at the boundaries and
thus stabilizing them. The primary determining factors, which affect wave dynamics
and transitions from one sort of dynamics to another, are
the conductances that govern the values of $G_{Kr}$ and $G_{CaL}$. 

We conclude that the chaotic dynamics of scroll waves is not only produced
by the common causes like a sharp APDR, EADs, and DADs, but also by a combined
variation of the rapid-rectifier and calcium currents, $I_{Kr}$ and
$I_{CaL}$; these play a crucial role in determining the dynamics of spiral 
and scroll waves in the two mammalian-heart models that we have studied. Our
detailed description of the dependence of spiral- and scroll-wave dynamics 
on changes in these currents should provide insights into an understanding 
of the effects of drugs that target these current channels. 

%
We mention some limitations of our study. We have used a monodomain model for
the cardiac tissue equations in our study; bidomain models are more realistic
than monodomain ones; however, a recent study~\cite{Potse} has shown that the
latter are adequate when currents are low, as in our study. To impose boundary
conditions we have used a phase-field approach~\cite{appndxphase}; this can also be done with a
finite-element model. The dynamics of scroll waves is affected by a many more parameters
than the two we study in detail. We have chosen these parameters for the reasons 
mentioned in this paper. A comprehensive study, including the simultaneous effects of 
change in more than two parameters, is computationally very expensive.

\section{Acknowledgments}

We thank  Council of Scientific and Industrial Research, University Grants
Commission and  Department of Science and Technology (India) for support, and
the Supercomputing Education and Research Centre (IISc) for computational
resources.


\end{document}